\documentclass[preprint,12pt]{elsarticle}

\usepackage{amssymb}
\usepackage{amsmath}
\usepackage{booktabs} 
\usepackage{hyperref}
\usepackage{adjustbox}
\usepackage{lineno}

\journal{Building and Environment}
\begin{document}

\begin{frontmatter}



\title{Effect of local and upwind stratification on flow and dispersion inside and above a bi-dimensional street canyon}


\author[author]{Davide Marucci}
\author[author]{Matteo Carpentieri\corref{mycorrespondingauthor}}
\ead{m.carpentieri@surrey.ac.uk}
\address[author]{EnFlo Laboratory, Department of Mechanical Engineering Sciences, University of Surrey, Guildford, Surrey GU2 7XH, UK}
\cortext[mycorrespondingauthor]{Corresponding author}
\begin{abstract}

The effects of a stably-stratified boundary layer on flow and dispersion in a bi-dimensional street canyon with unity aspect ratio have been investigated experimentally in a wind tunnel in combination with differential wall heating. Laser-Doppler anemometry together with a fast flame ionisation detector and cold-wire anemometry were employed to sample velocities, concentration, temperatures and fluxes.

A single-vortex pattern was observed in the isothermal case, preserved also when leeward wall was heated, but with a considerable increment of the vortex speed. Heating the windward wall, instead, was found to generate a counter-rotating vortex, resulting in the reduction of velocity within the canopy. The stable stratification also contributes reducing the speed, but only in the lower half of the canyon. The largest values of turbulent kinetic energy were observed above the canopy, while inside they were concentrated close to the windward wall, even when the leeward one was heated. An incoming stable stratification produced a significant and generalised turbulence reduction in all the cases. Windward heating was found to produce larger temperature increments within the canopy, while in the leeward case heat was immediately vacated above the canopy. A stable approaching flow reduced both the temperature and the heat fluxes.

A passive tracer was released from a point source located at ground level at the centre of the street canyon. The resulting plume cross-section pattern was mostly affected by the windward wall heating, which produced an increment of the pollutant concentration on the windward side by breaking the main vortex circulation. The application of an incoming stable stratification created a generalised increment of pollutant within the canopy, with concentrations twice as large. Turbulent pollutant fluxes were found significant only at roof level and close to the source. On the other hand, in the windward wall-heated case the reduction of the mean flux renders the turbulent component relevant in other locations as well.

The present work highlights the importance of boundary layer stratification and local heating, both capable of creating significant modifications in the flow and pollutant fields at microscale range.
\end{abstract}

\begin{keyword}
Differential heating \sep
Stable boundary layer \sep
Wind tunnel \sep
Street canyon \sep
Dispersion


\end{keyword}

\end{frontmatter}


\section{Introduction}
\label{Introduction}

Due to rapid urbanisation, air pollution in the urban environment is an increasing problem, especially in developing countries. Together with ordinary exposure to pollution, another threat to the human health is represented by incidents involving the release in the atmosphere of toxic gases or radioactive substances. The capacity of predicting gas and particle dispersion can assist in preventing health hazards and planning emergency procedures. However, one of the main problems affecting this kind of models is the way they treat thermal stratification, very often present in environmental flows (see e.g. \cite{Wood2010} for field observations over the city of London, UK). Atmospheric stratification involves differences in air density caused by a positive (stable) or negative (unstable) vertical gradient of virtual potential temperature. The stability of the layer depends on the stratification and affects the atmospheric boundary layer depth and structure as well as velocity, temperature and turbulence properties. On the other hand, buoyancy effects on the flow may also be caused by local sources of heating (e.g. differential heating of building walls or ground due to solar radiation or human activity). At the microscale range both of these effects may be significant and are worth to be investigated.

One of the most interesting (and hence most studied) urban geometric unit is represented by the street canyon. Some field studies have been performed so far. \citet{Nakamura1988} found that the largest increase in temperature was confined within 0.5~m of the floor or walls with the canyon air remaining thermally unstable also during night in a hot Summer. Differently, \citet{Niachou2008} observed an inversion of 7$^\circ$C$/$100m  at the centre of a street canyon in Athens during the morning. \citet{Louka2002} found temperature gradients up to 10$^\circ$C in the vicinity of the sun-heated walls in Nantes.

Compared to field measurements, only few wind tunnel studies have been attempted to date, mainly focussing on the effects of buoyancy forces. \citet{Uehara2000} simulated an array of aligned cubic blocks with stratified (stable and convective) approaching flow. On the other hand, \citet{Kovar-Panskus2002} and \citet{Allegrini2013} focussed on local stratification, investigating the case of differential heating for a street cavity. While the former only studied the case of windward wall heating, the latter extended their study to cases where either the leeward, ground or all three surfaces were heated. \citet{Kovar-Panskus2002} only found a weak secondary vortex arising when the windward wall was heated, while \citet{Allegrini2013} noted a clear counter-rotating vortex. When the leeward wall was heated, on the contrary, the flow structure remained unaltered compared to the isothermal case, with only an increment of the mean velocity and the turbulent kinetic energy (TKE). 3D canyons with different building length and roof shape subject to ground heating were, instead, considered by \citet{Allegrini2018}. The heated case presented completely different horizontal flow patterns along the canyon axis, which affected also the vertical one, destroying the typical single vortex structure observed in the isothermal case with unity aspect ratio \cite{Oke1988}. Their results highlight how buoyancy can affect the three-dimensionality of street-canyon flows.

The literature about numerical simulations is wider and more diverse. \citet{Sini1996} was among the first to numerically demonstrate the influence of thermal forcing on pollutant dispersion in street canyons with Reynolds-averaged Navier–Stokes (RANS) simulations. \citet{Kim1999} and \citet{Xie2005, Xie2007} further investigated cases with differential wall heating, different canyon aspect ratio and building height. These works highlighted how the mean flow pattern in street canyons can be modified by both geometric and thermal factors, with the main vortex structure strengthened, weakened or broken into multiple vortices as effect of canyon surface heating. Various canyon geometry configurations were studied by \citet{Mei2018, Mei2019}, who focussed on groups of street canyons ventilated merely by thermal buoyancy force induced by uniformly heating the building surfaces. They considered sets of increasing number of 2D canyons by sldo investigating different aspect ratios ($H/W$ 0.5 to 3) and building heights, either alternating taller and smaller buildings or assuming rising or reducing heights throughout the sets. Thermal plumes were found to converge, resulting in a stagnant region at the urban centre with a peak value of the temperature.

The widespread diffusion of the use of Large-Eddy Simulations (LES) has helped in analysing more in detail the modifications in the turbulence structure. \citet{Li2010,Li2012} considered the ground heating case and varying aspect ratio ($H/W = 0.5$, 1 and 2). In their results, buoyancy increased the flow velocity, turbulence and turbulent pollutant flux inside the canopy, as well as the pollutant removal. On the other hand, the shear layer at roof level appeared weakened. \citet{Li2016} extended their investigation to the case of stable boundary layers, too, by means of cooling the ground. In this case velocities were reduced by the buoyancy, the turbulent pollutant flux close to the leeward wall became negative and pollutant was trapped in the lower region of the canopy. Also \citet{Cheng2011a} analysed a similar case with heated and cooled ground obtaining conclusions quite similar.
Cai \cite{Cai2012,Cai2012a} investigated cases with differential wall heating: either the leeward wall or the windward wall were heated together with the building roof, while pollutant was released from either ground or canyon wall surfaces. When the leeward wall was heated, the mean flow pattern was approximately symmetric while the main vortex extended to heights above the roof level and was accelerated. On the other hand, windward heating generated an asymmetric pattern with velocity clearly suppressed, accompanied by an increase of TKE. The turbulent pollutant fluxes were significant only at roof level and above the canopy in the leeward-heated case while for the windward case they were comparable with the advective fluxes inside the canopy and predominant above. Recently there have also been attempts to numerically investigate realistic wall heating patterns in three-dimensional urban configurations \cite{Nazarian2018}, as opposed to uniformly heated surfaces. In this regard, \citet{Nazarian2018} stressed the importance of considering a detailed three-dimensional heating for studies of thermal comfort. In case the concentration field is of interest, instead, they found it mainly affected by the overall heating of the surfaces, while a detailed three-dimensional heating was deemed superfluous.

Analysing the available literature we came to the conclusion that there is a shortage of experimental data dealing with stratified flows problems in street canyons. Moreover  none of the mentioned experiments included dispersion measurements. For this reason, an experimental investigation has been undertaken at the EnFlo laboratory. Initially it focussed on improving the technique to accurately simulate stratified (both stable, SBL, and convective, CBL) boundary layers in the wind tunnel, suitable for high roughness surface conditions. The results were reported by \citet{Marucci2018}. Then, the generated boundary layers were applied as approaching flows to an array of rectangular blocks and turbulent pollutant and heat fluxes were measured \cite{Marucci2018a}. Clear effects on the plume height and concentration levels were observed from a ground level source release, in that study.

Here, the case of an isolated bi-dimensional street canyon with unity aspect ratio is considered. Five heating configurations were investigated during the experiments, but only three are reported here for brevity and because they were the most interesting ones: no heating (NH), windward wall heated (WH) and leeward wall heated (LH). The other two configurations not shown here are ground-heated and all surfaces-heated. The measurements are repeated with neutral (NBL) and stable approaching boundary layers (indicated as SNH, SWH and SLH, respectively for the three cases highlighted above) to investigate the combined effects of approaching flow and local stratification. To the knowledge of the authors, this represents an absolute novelty in the literature of urban ventilation.

\section{Methodology}
\label{Methodology}
\subsection{Wind tunnel and flow generation}
The experiments were carried out in the EnFlo meteorological wind tunnel, at the University of Surrey. The open-return facility is characterised by a working section 20~m long, 3.5~m wide and 1.5~m high. A set of seven Irwin's spires \cite{Irwin1981} was employed to artificially thicken the boundary layer. They were 986~mm high, 121~mm wide at the base and 4~mm at the tip, laterally spaced 500~mm, specifically developed to generate a SBL about 850~mm deep ($\delta$) \cite{Marucci2018}. Rectangular-shaped sharp-edge roughness elements were also placed on the floor in a staggered arrangement, 240~mm apart laterally and 240~mm spaced streamwise. This was to guarantee the development of a rough approaching flow for the model. When a SBL was simulated, a vertical temperature profile was imposed at the inlet section by means of a series of fifteen 100~mm-high horizontal heaters while a negative surface heat flux was generated with floor-cooling panels by means of recirculating water. The same water was also employed to keep the laboratory at a constant temperature by cooling the air leaving the wind tunnel. Floor temperature was measured with thermistors attached to the floor every 2~m and averaged together. Temperature variations within $\pm 0.3^\circ$C were observed but deemed acceptable.  For more details see \cite{Marucci2018}.

\begin{figure*}
	\centering
	\includegraphics[width=\textwidth]{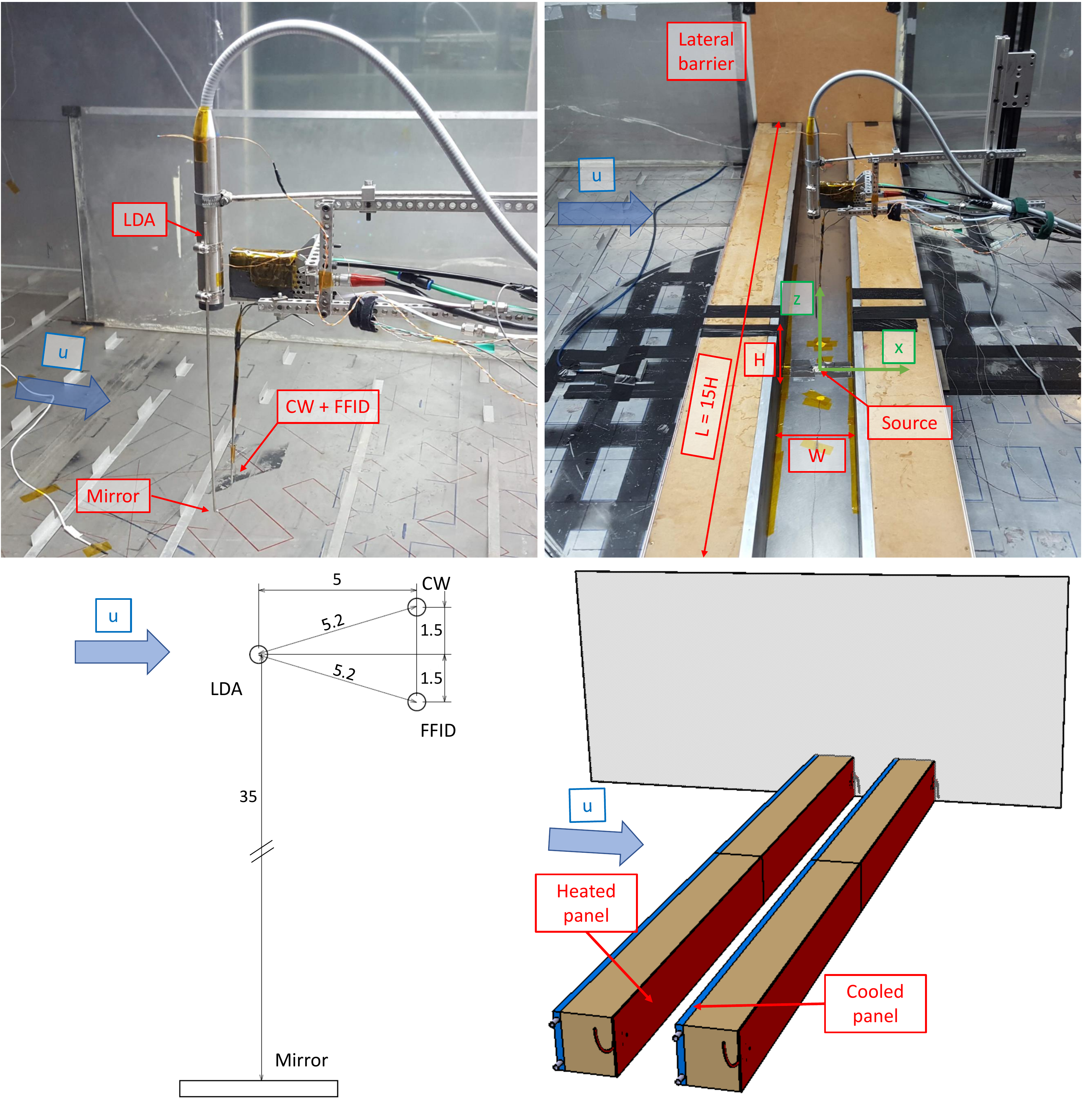}
	\caption{Photo and scheme of the measuring setup (on the left) and model geometry (on the right)}
	\label{fig:ModelSetup}
\end{figure*}

\subsection{Street canyon model}
Photo and scheme of the model geometry are displayed in Figure~\ref{fig:ModelSetup}.
The model geometry was an isolated bi-dimensional street canyon, with aspect ratio ($H/W$) equal to 1 and length-to-height ratio ($L/H$) of 15. The model height ($H$) was 166~mm (about $1/5$ of the approaching boundary layer depth $\delta$), chosen as a compromise between the necessity of minimising the blockage (here 7.9\%) and providing at the same time sufficiently large wall surfaces for heat exchange when dealing with local heating. The square cross-section buildings were designed in order to have one wall heated with electrical heater mats (with power 4 kW$/$m$^2$) and the opposite one cooled by means of circulating water (the same used for the wind tunnel floor). The experiments were repeated with neutral and stable approaching flow in order to evaluate the combined effect of incoming and local stratification. A mixture of air and propane was released at ground level from a circular source with a diameter of 22~mm at the centre of the street canyon. The hole was filled with plastic beads and the mixture emission velocity was maintained equal to $0.03U_{REF}$ (with $U_{REF}$ reference velocity later defined) in order to guarantee a passive emission.
Lateral barriers where added to increase the flow bi-dimensionality.

In the results presented below, the origin of the reference system is the centre of the street canyon, placed at 14~m from the working-section inlet. $z$ is the distance from the wind tunnel floor; $y$ is aligned with the street canyon centreline. $\overline{U}$, $\overline{V}$ and $\overline{W}$ represent the time-averaged velocity on the $x$, $y$ and $z$ directions, respectively, while $u'(t)$, $v'(t)$ and $w'(t)$ are the fluctuations (e.g. for the streamwise component $u(t)=\overline{U}+u'(t)$). $U_{REF}$ is a reference velocity measured with a sonic anemometer at 5~m from the inlet ($y=1$~m, $z=1$~m).

\subsection{Measuring setup}
\begin{figure}
	\centering
	\includegraphics[width=\linewidth]{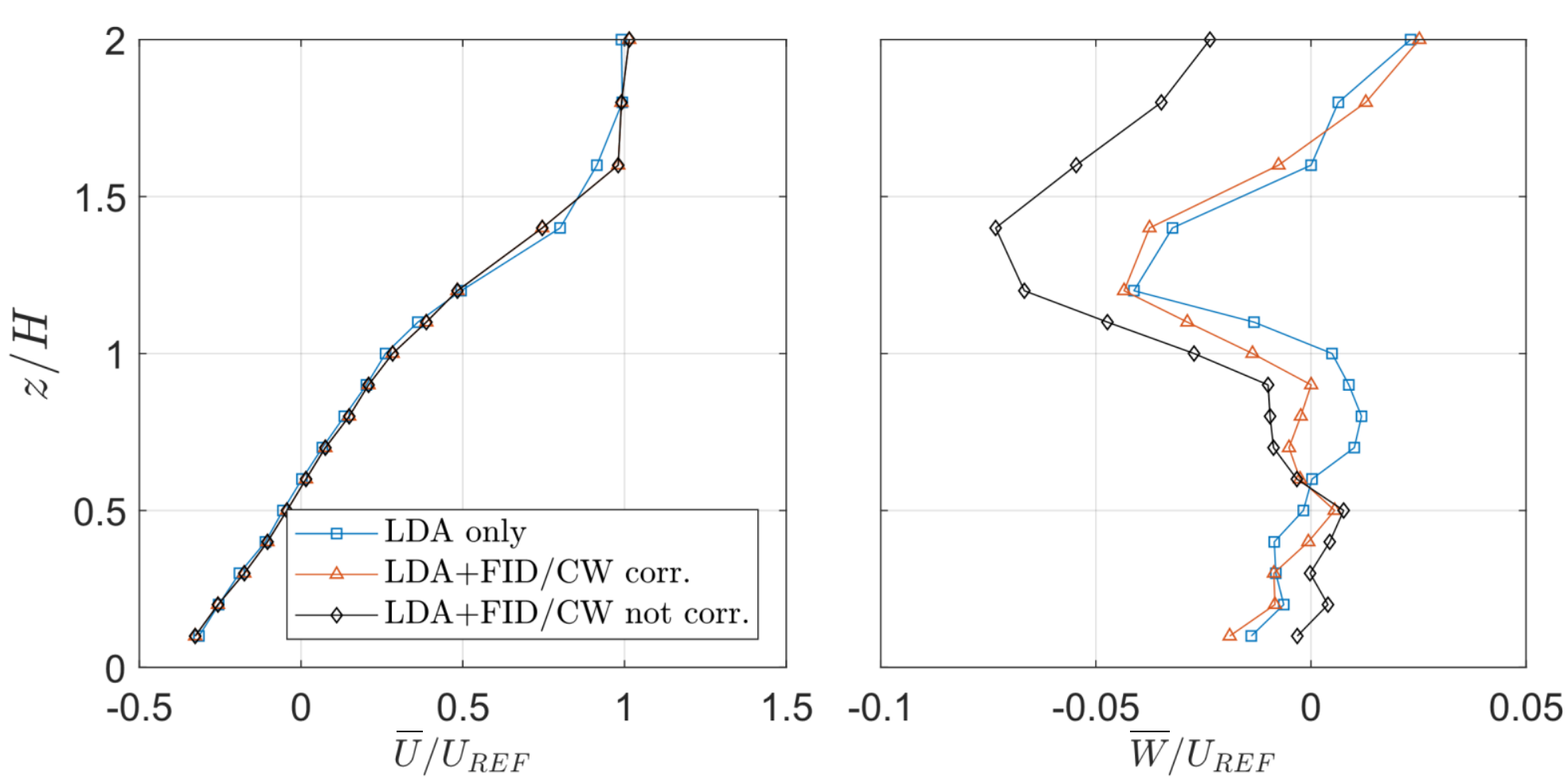}
	\caption{Effect of the FID/CW presence on the vertical profiles of mean streamwise and vertical velocity at $x/H = 0$. Also the corrected profile is shown.}
	\label{fig:VelCorrection}
\end{figure}

Figure~\ref{fig:ModelSetup} also presents photo and scheme of the measuring setup. Velocity measurements were performed by means of a two-component laser-Doppler anemometer (LDA), via a Dantec 27~mm FibreFlow probe. The target acquisition frequency was set to 100~Hz, while the LDA focal length was 300~mm. A small mirror was added 35~mm on one side to deflect the laser beams and measure the vertical component of the velocity. Fluctuating temperatures and concentrations were sampled at 1000~Hz using a calibrated fast-response cold-wire probe (CW) and a fast flame ionisation detector system (FFID), held close to each other and placed 5~mm downstream the LDA measuring volume, so as to allow measurement of the turbulent heat and pollutant fluxes. Since velocity, temperature and concentration measurements took place at the same time, the LDA location was used as the main reference for the measurement position, while CW and FFID were measuring 5~mm downstream. The actual position of the probes is shown in the in the mean temperature and concentration plots, while the reference LDA position was used when plotting fluxes. The presence of the FFID and CW probes was found to produce a small perturbation on the mean vertical velocity measurement, while no significant effects were identified for the mean streamwise component, as well as in variances and covariances. To minimise the observed bias the following correction was applied to the vertical mean value: $\overline{W}_{corr}=\overline{W}+\overline{U}\sin{\left(2.75^\circ\right)}$ (an example of the correction is shown in Figure~\ref{fig:VelCorrection}). The effect of the temperature on the FFID was found negligible. Nevertheless, the system was calibrated every two hours during the measurements, while the background level of concentration was monitored every 20 minutes and subtracted to the measured value. The delay time occurring between the LDA and FFID signal (mainly due to the FFID sampling tube length) was evaluated by analysing the cross-correlation of pollutant and streamwise velocity immersed in a jet of polluted air (a delay time in the range between 15 and 30~ms was applied).  Moreover, the dilution ratio of propane in the mixture of the tracer gas was adjusted in the range 0.5-1.8\% to keep the sampled concentration within the dynamic range of the FFID.

\subsection{Measuring error estimation and flow bi-dimensionality}
The standard error for first and second order statistics was evaluated for each measuring point. On average, the standard error for the first order statistics of velocity and concentration is below $\pm10\%$. For the temperature and velocity variance it is about 7\%, while for the concentration variance it is larger (23\%). Finally, for the covariances of velocity and temperature it is of the order of 20\%, again larger for the concentration covariances (30\%). The high value observed for mean velocities, compared to the variance, is mainly due to the fact that in many points velocities are very close to zero. For this reason, points with error larger than 150\% have been filtered out in the average calculations.

The lateral variability of flow quantities was also investigated in order to assess the bi-dimensionality. Two lateral profiles at $x/H = -0.3$, $z/H = 0.2$ and $x/H = 0.3$, $z/H = 0.9$ have been measured in the range $y/H = \pm 3$ for each case. On average, the streamwise velocity variability was in the range $\pm 18\%$ compared to the mean value and $\pm 10\%$ for the vertical component. The temperature was laterally quite uniform ($\pm 1\%$). Velocity and temperature variances were within 15\%, more variability for the covariances ($\pm 50$ and 30\% for velocity and temperature, respectively). Overall, the uniformity in the investigated range was deemed satisfactory. Finally, it is worth mentioning that for all the contour graphs and spatially-averaged statistics displayed in the following paragraphs, experimental data have been interpolated by using the ``natural neighbour method'' \citep{Sibson1981} on a grid with resolution $H/100$.

\subsection{Approaching flow}
Two different types of approaching flows were studied, a neutral and a stable boundary layer. The scaling characteristics of the two boundary layers are reported in Table~\ref{table:BL}. $U_{REF}$ was chosen equal to 0.65~m$/$s. This quite low velocity was necessary to obtain appreciable local stratification effects within the canyon. The Reynolds numbers based on the length and velocity at the boundary layer top ($Re_\delta$) and building roof ($Re_H$) were about $4\times10^4$ and $5\times10^3$, respectively. A more detailed discussion on the Reynolds number independence issue is reported in section \ref{Reynolds}. The boundary layer depth was approximately equal to 5 times the model height. $\Delta\Theta$ is the difference between the air temperature at the boundary layer top ($\Theta_{\delta}$) and the wind tunnel floor temperature ($\Theta_0$). The friction velocity $u_\ast$ was evaluated as $u_\ast = \sqrt{-\left(\overline{u'w'}\right)_0}$, in which $\left(\overline{u'w'}\right)_0$ was extrapolated as linear fitting to the floor from the data. Similarly, the friction temperature was computed as $\theta_\ast = -\left(\overline{w'\theta'}\right)_0/u_\ast$ for the stable case.
The aerodynamic roughness length ($z_0$) was evaluated with a non-linear fitting of the equation

\begin{equation} \label{u_mo_SBL}
\overline{U}(z) = \frac{u_\ast}{k}\left[\ln \left(z/z_0\right)+8\frac{z-z_0}{L_{O}}\right]
\end{equation}

\noindent using the mean streamwise velocity profile (see \cite{Marucci2018}). $k$ is the von Karman constant (assumed here equal to 0.40) while the displacement height was found approximately equal to zero (hence not reported in the equation). The Monin-Obukhov length $L_{O}$ is expressed as
\begin{equation}
L_{O} =-\frac{\Theta_0}{kg}\frac{u_\ast^2}{\theta_\ast}
\end{equation}
\noindent where $g$ is the acceleration of gravity. Similarly, the thermal roughness length ($z_{0h}$) was estimated by means of fitting

\begin{equation} \label{t_mo_SBL}
\overline{\Theta}(z) =  \Theta_0 + \frac{\theta_\ast}{k}\left[\ln \left(\frac{z}{z_{0h}}\right)+16\frac{z-z_{0h}}{L_{O}}\right]
\end{equation}

\noindent using the mean temperature profile (see \cite{Marucci2018}).

Three non-dimensional numbers are given to quantify the approaching flow stability level. The ratio $\delta/L_{O}$ and the bulk Richardson number, evaluated at the boundary layer top $Ri_\delta$ and at model top $Ri_H$.

\begin{equation} \label{bulkRi}
Ri_\delta = \frac{g\left(\Theta_\delta - \Theta_0\right)\delta}{\Theta_0 U_\delta^2},\ \
Ri_H = \frac{g\left(\Theta_H - \Theta_0\right)H}{\Theta_0 U_H^2}
\end{equation}

Finally, also the roughness Reynolds number $Re_\ast = z_0u_\ast/\nu$ (the kinematic viscosity $\nu$ is the one at floor temperature for all three Reynolds numbers evaluated), was calculated.

\begin{figure*}
	\centering
	\includegraphics[width=0.7\textwidth]{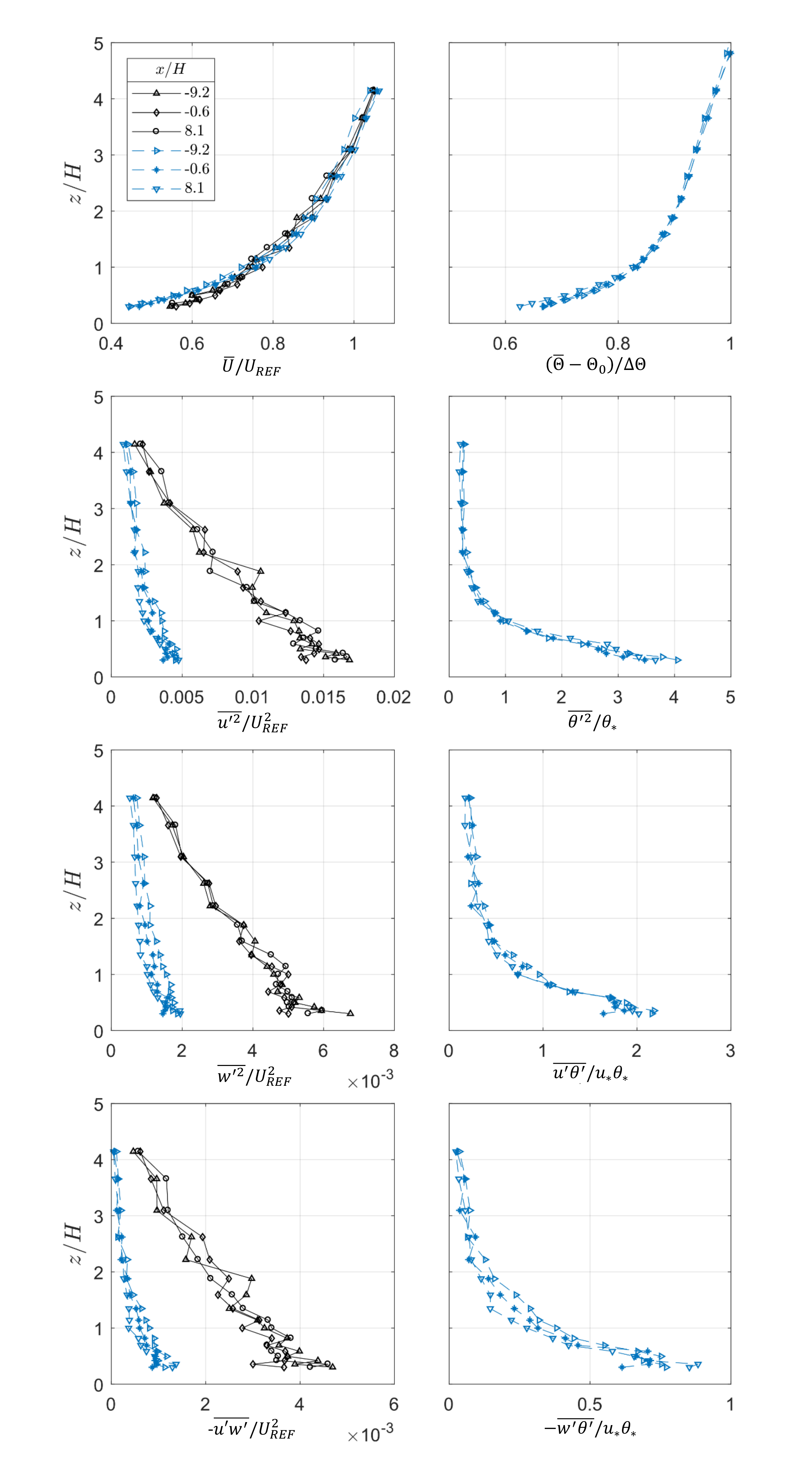}
	\caption{First and second order statistics for the approaching flow. Black lines are NBL while blue are SBL.}
	\label{fig:BL}
\end{figure*}

Vertical profiles of first and second order statistics of velocity and temperature are displayed in Figure~\ref{fig:BL} for three locations along the wind tunnel centreline, acquired without the street canyon model. The most evident effect of the stable stratification on the approaching flow is the large dampening in the turbulence, well represented by the friction velocity reduction of almost 50\%. Differently, the mean velocity profile is only slightly modified, according to what observed also by \citet{Marucci2018}, to which we refer for further comments.

\begin{table}
	\caption{Main scale parameters for neutral and stable approaching flow}
	\centering
	\label{table:BL}
	\begin{tabular}{l c c}
		\toprule
		 & NBL & SBL \\
		\midrule
		$U_{REF}\ \mathrm{(m/s)}$ & 0.65 & 0.65 \\
		$\delta/H$ & $\approx 5$ & $\approx 5$ \\
		$\Theta_0\ \mathrm{(^\circ C)}$ & 24 & 19 \\
		$\Delta\Theta\ \mathrm{(^\circ C)}$ & 0 & 7 \\
		$u_\ast/U_{\text{REF}}$ &0.065 & 0.035\\
		$\theta_\ast\ \mathrm{(K)}$ &-& 0.12\\
		$z_0\ \mathrm{(mm)}$ & 1.6 & 1.2\\
		$z_{0h}\ \mathrm{(mm)}$ & - & $\approx 0.001$\\
		$\delta/L_{O}$ & 0 & 2.7 \\
		$Ri_\delta$ & 0 & 0.39 \\
		$Ri_H$ & 0 & 0.13 \\
		$Re_\delta\ (\times10^3)$ & 37.8 & 40.5 \\
		$Re_H\ (\times10^3)$ & 5.2 & 5.3 \\
		$Re_\ast$ & 4.7 & 1.8 \\
		\bottomrule
	\end{tabular}
\end{table}

\subsection{Reynolds number effect}
\label{Reynolds}
\begin{figure}
	\centering
	\includegraphics[width=\linewidth]{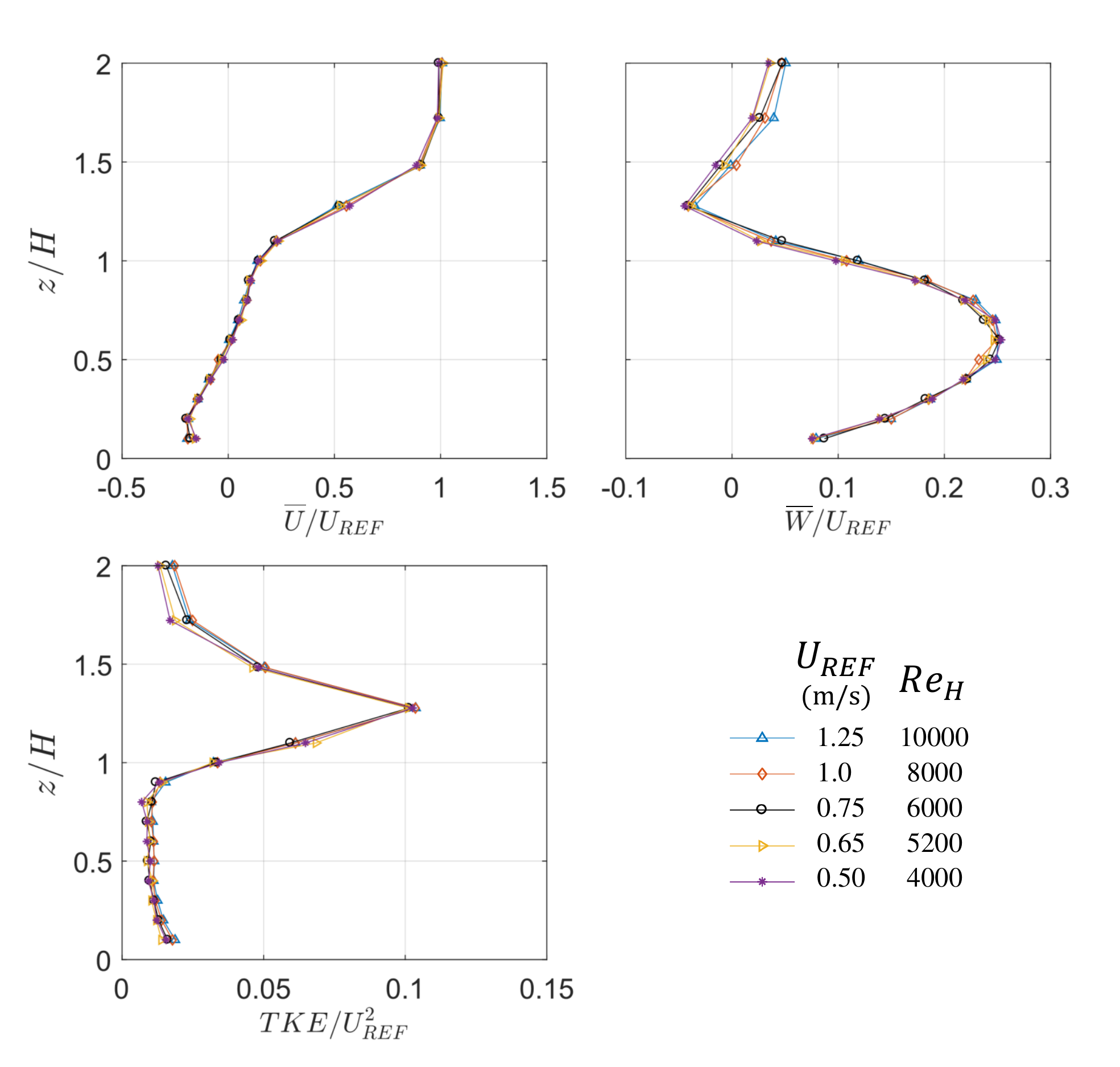}
	\caption{Mean streamwise, vertical velocity and TKE for different reference velocities, equivalent to $Re_H$ 10000, 8000, 6000, 5200, 4000. $x/H$ = -0.3.}
	\label{fig:Re1}
\end{figure}

\begin{figure}
	\centering
	\includegraphics[width=\linewidth]{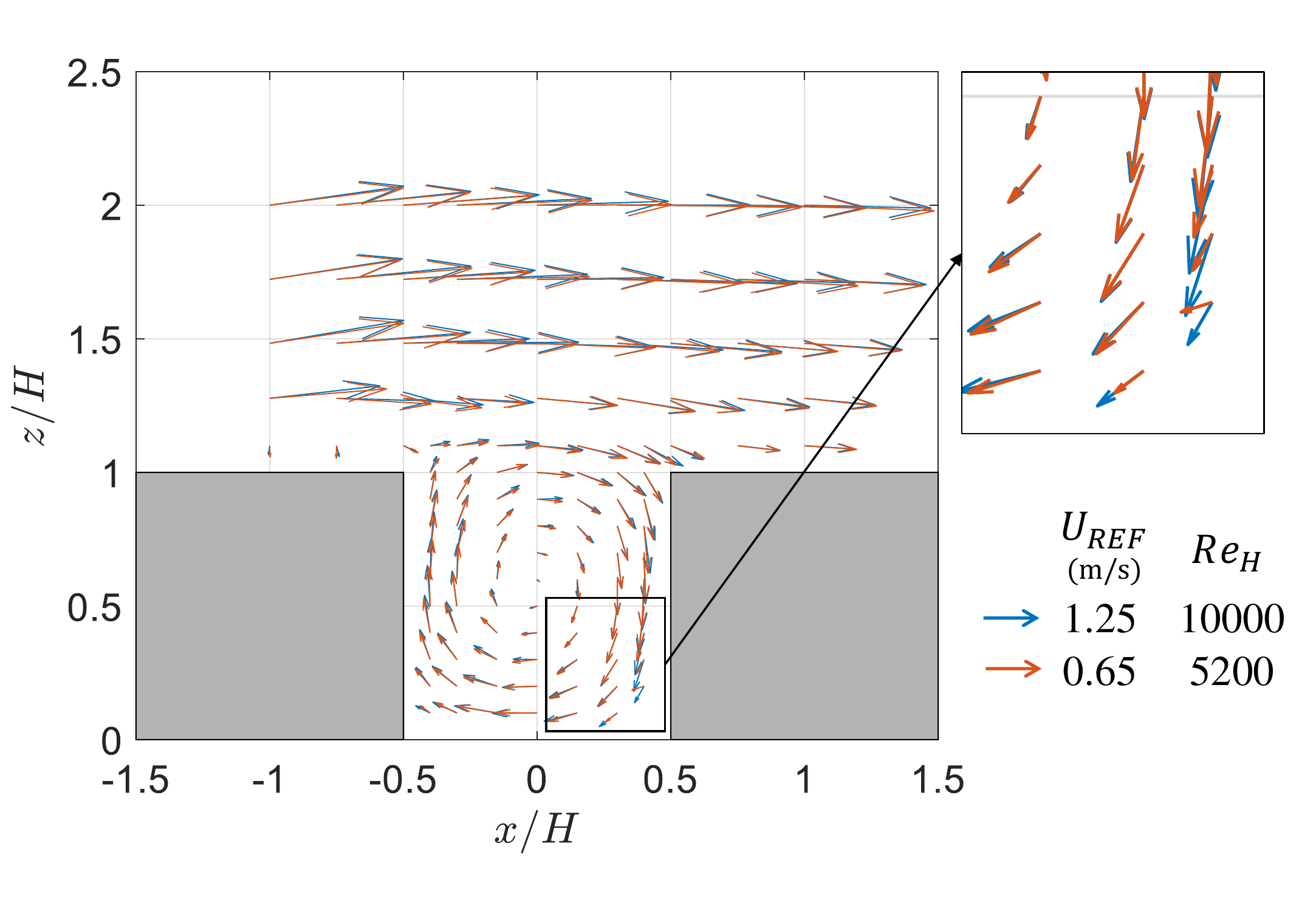}
	\caption{Flow velocity vectors for two Reynolds numbers ($Re_H$ = 5200, 10000).}
	\label{fig:Re2}
\end{figure}

Reynolds number independence is a key feature of fluid dynamics experiments to guarantee that normalised velocities are representative of the full-scale flow field. The necessity to work with small velocities to obtain reasonable buoyancy effects with reasonable wall temperatures in local stratification studies means that Reynolds independence might be difficult to satisfy. In order to assess the Reynolds number effect for the chosen velocity the isothermal case was repeated with different reference speeds (varying from 0.5 to 1.25~m/s). Figure~\ref{fig:Re1} shows a vertical profile of the mean velocities and TKE. The TKE is evaluated as $3/4\left(\overline{u'^2}+\overline{w'^2}\right)$, assuming that the lateral component ($\overline{v'^2}$, not measured) behaves like the average of the other two \cite{Allegrini2013}. The measurements show that $\overline{U}$ is rather insensitive to the Reynolds number in that range, while $\overline{W}$ experiences a slight reduction above the canopy for the two lower velocities considered. The same can be said for the TKE which, in the $U_{REF}=0.65$~m$/$s case, sees an average reduction of 5\% above the canopy and 9\% within it, compared to the 1.25~m$/$s case. These can be considered small and we can reasonably take the $U_{REF}=0.65$~m/s case as representative for a full-scale flow. Figure~\ref{fig:Re2} shows the velocity vectors for the 1.25 and 0.65 cases. The most critical part is represented by the canyon lower-right corner (also visualised in the magnified window). Here the lower velocity case appears to differ the most, but the region affected is also quite limited in space, so that it does not seem to affect a large portion of the flow field.

Finally, it is worth mentioning that in all the cases presented here, $Re_H$ is always larger than 3400, which is the critical value indicated by \citet{Hoydysh1974} to have independence from viscous effects in the street canyon flow pattern. The result is also supported by the fact that $Re_\ast$ (used to evaluate whether the surface is fully rough) is, for the slowest case, still greater than 1, which is the minimum value indicated by \citet{Snyder2002} for sharp-edged roughness elements in a NBL. It is important to highlight, though, that all Reynolds number independence discussion reported here is referring to the isothermal case. Very recently \citet{Chew2018}, after performing LES simulations at different scales, pointed out that a non-isothermal case may not be Reynolds number independent, even though the isothermal is, and suggested to be careful in extending conclusions obtained with reduced-scale model to the full-scale case if buoyancy forces are considered. Further studies will have to be conducted to address this point, even though the investigation of a meaningful range of Reynolds numbers can prove to be very challenging in stratified wind tunnels.

\section{Results}
\label{Results}
Table~\ref{table:ModelStrat} lists the local scaling quantities for the different experimental cases, which will be used to normalise the graphs in the following paragraphs.

\begin{table}
	\caption{Local scaling quantities for the street canyon}
	\centering
	\label{table:ModelStrat}
	\begin{adjustbox}{width=\columnwidth,center}
		\begin{tabular}{l c c c c c c}
			\toprule
			& NH & WH & LH & SNH & SWH & SLH \\
			\midrule
			$U_{2H}\ \mathrm{(m/s)}$ & 0.66 & 0.64 & 0.66 & 0.65 & 0.66 & 0.65 \\
			$\Theta_{2H}\ \mathrm{(^\circ C)}$ & 24.0 & 24.2 & 23.9 & 25.1 & 25.1 & 25.1 \\
			$\Theta_{GROUND}\ \mathrm{(^\circ C)}$ & 24.0 & 23.5 & 25.5 & 19.7 & 21.4 & 22.1 \\
			$\Theta_{HOT}\ \mathrm{(^\circ C)}$ & - & 118.5 & 120 & - & 118.0 & 118.0 \\
			$Ri_{Local}$ & - & -1.27 & -1.22 & - & -1.18 & -1.19 \\
			$Fr_{Local}$ & - & -0.79 & -0.82 & - & -0.85 & -0.84 \\
			\bottomrule
		\end{tabular}
	\end{adjustbox}
\end{table}

$U_{2H}$ and $\Theta_{2H}$ are, respectively, the mean streamwise velocity and temperature measured at $x/H=0$, $z/H=2$. They will also be used to normalise the respective quantities in the following graphs, so that a comparison with the literature (widely using a similar scaling) is possible. Nevertheless, whenever relevant, other normalisations will be considered. $\Theta_{GROUND}$ is the temperature of the ground measured inside the street canyon, while $\Theta_{HOT}$ is the temperature of the heated building wall. A local Richardson number is defined to quantify the local stratification in case differential wall heating is applied. It is defined as

\begin{equation} \label{bulkRiLoc}
Ri_{Local} = \frac{g\left(\Theta_{2H} - \Theta_{HOT}\right)H}{\Theta_{2H} U_{2H}^2}
\end{equation}

\noindent For completeness also the Froude number is indicated in the table (equivalent to $Fr_{Local} = Ri_{Local}^{-1}$).

\subsection{Flow and turbulence} \label{Sec:Flow and turbulence}

\begin{figure*}
	\centering
	\includegraphics[width=\textwidth]{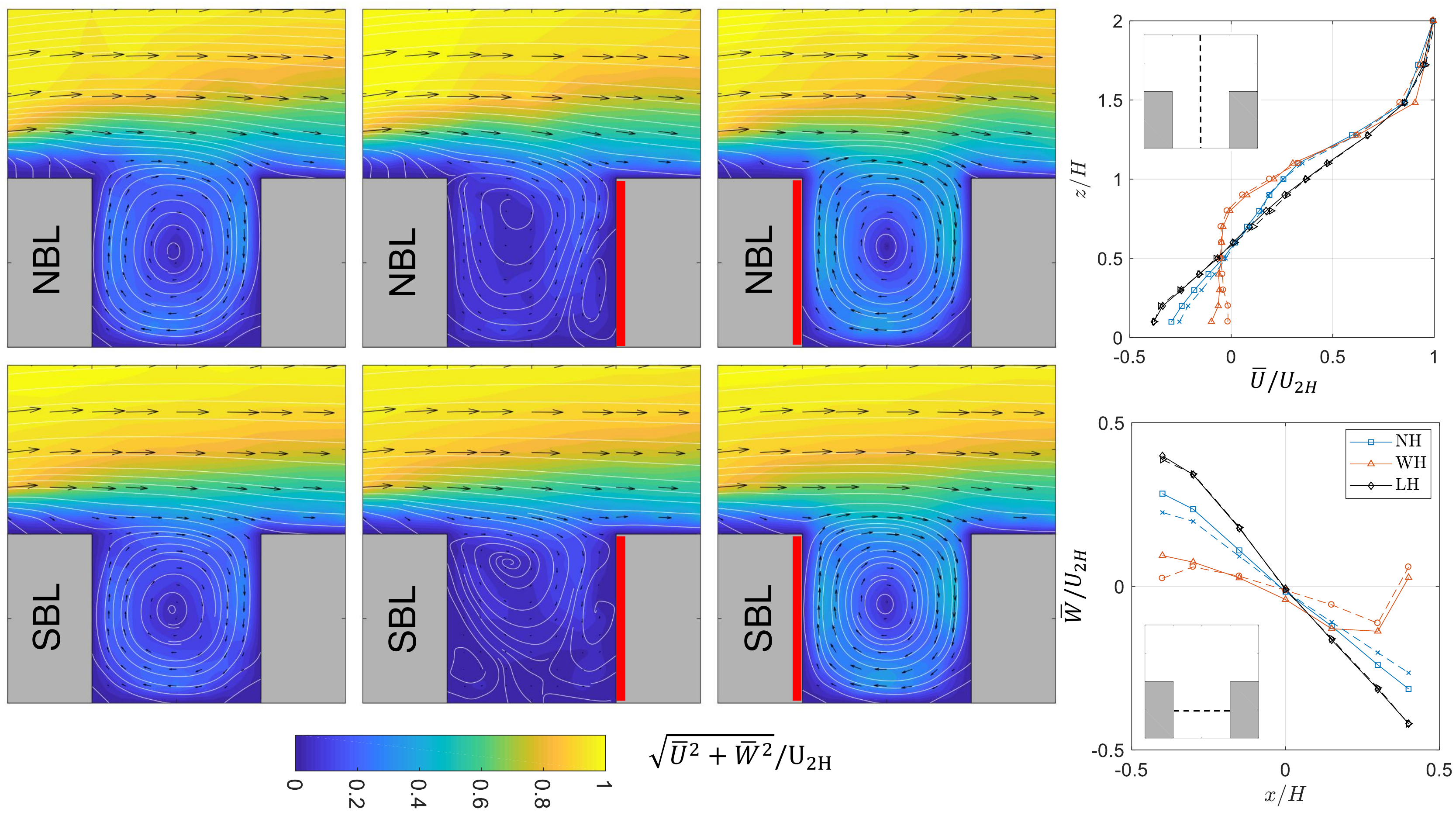}
	\caption{Contours and vectors of mean velocity, white lines are streamlines. The red lines represent the heated surfaces in each case ([S]NH left, [S]WH centre, [S]LH right). The line plots on the right show the vertical profiles of mean streamwise velocity at $x/H = 0$ (top) and the longitudinal profiles of mean vertical velocity at $z/H = 0.5$ (bottom); NBL = continuous lines, SBL = dashed lines.}
	\label{fig:Vel1}
\end{figure*}

\begin{figure*}
	\centering
	\includegraphics[width=\textwidth]{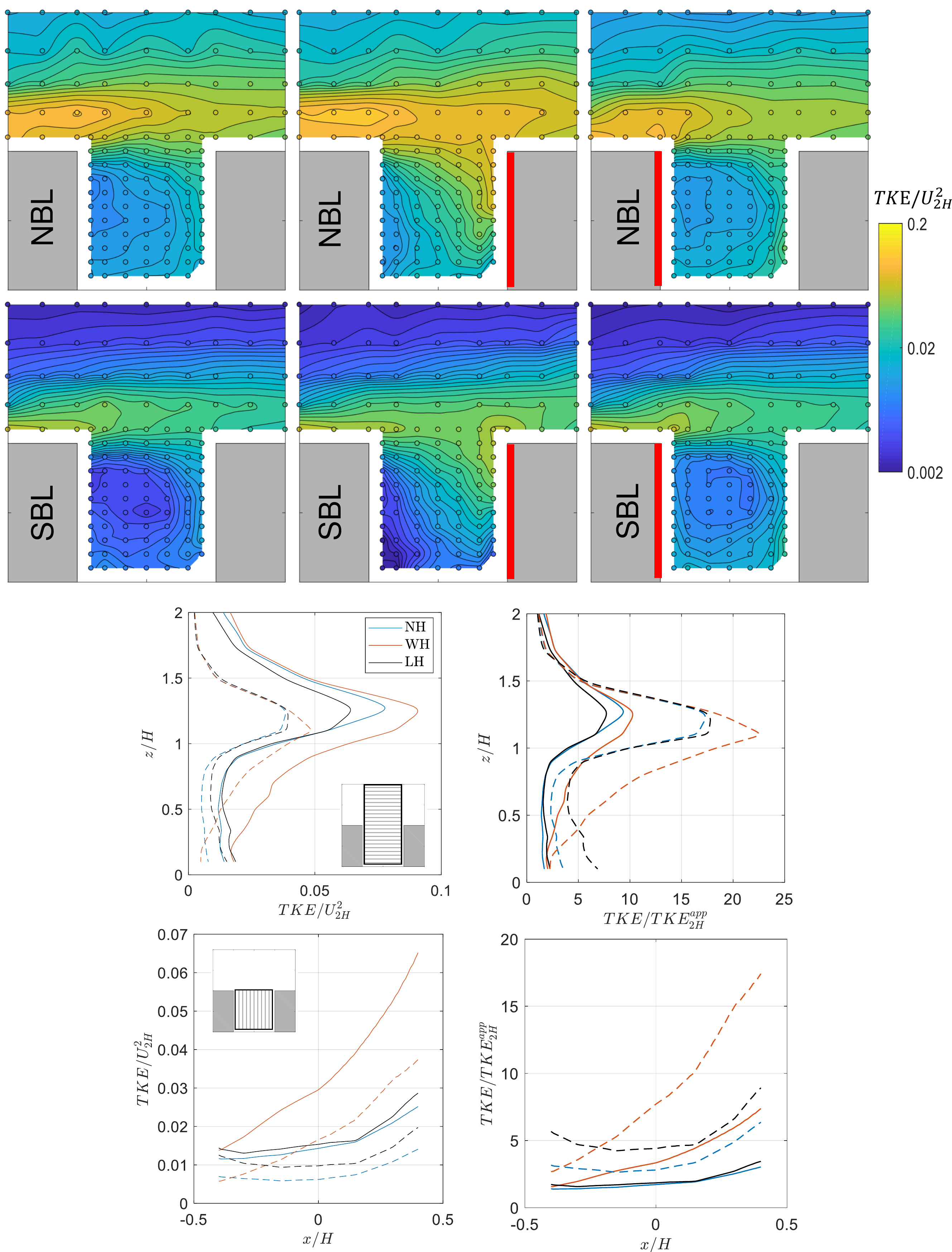}
	\caption{Contours of TKE. The red lines represent the heated surfaces in each case ([S]NH left, [S]WH centre, [S]LH right). The line plots at the bottom show the vertical profiles of longitudinally-averaged TKE at $x/H = 0$ (top) and the longitudinal profiles of vertically-averaged TKE at $z/H = 0.5$ (bottom), normalised by the reference velocity (left) or the approaching flow TKE at $z = 2H$ (right); NBL = continuous lines, SBL = dashed lines.}
	\label{fig:TKE1}
\end{figure*}

Figure~\ref{fig:Vel1} shows the contours of normalised mean velocity, as well as velocity streamlines. Also vertical profiles for $x/H = 0$ and longitudinal profiles at $z/H = 0.5$ are presented for the various configurations. The flow structure inside the canyon when no local heating was applied and an incoming NBL is characterised by a single-vortex pattern whose centre is located at $x/H$ = 0 and approximately at a height of $z/H = 0.6$. Differently from what other authors suggested (e.g., \cite{Allegrini2013}, \cite{Li2016} and \cite{Park2012}, no secondary vortices are present close to the bottom corners, but this may be due to the lower resolution of the measurement grid (the closest measuring point to the surfaces is $0.1H$ from them). The above structure is present with only minor modifications in the LH case as well. Differently, in the WH case a second counter-rotating vortex arises, generated by the buoyancy forces produced by the heated wall, which opposes the descending motion of the air into the canyon, hence slowing down the velocity (as better shown by the profiles on the right-hand side of the figure). A similar behaviour was observed by several authors (e.g. \cite{Sini1996}, \cite{Allegrini2013}, \cite{Cai2012}), hence a consensus seems to have been established. The centre of the main vortex appears shifted toward the upper corner of the leeward building and, on average, the mean velocity within the canyon is 50\% lower than in the NH case. In the LH case, on the other hand, the buoyancy forces act accelerating the flow, thus resulting in a 37\% average increment of the velocity within the canyon.

The application of a stable approaching flow has an evident effect on reducing the mean velocity, mainly in the bottom half of the canyon. \citet{Li2016} simulated a similar level of stability for the approaching flow in bi-dimensional street canyons and they too found similar conclusions. However, in their case this effect was more accentuated, bringing to the formation of real stagnation regions closer to the ground. In our measurements the reduction is more modest, but it should be stressed that the geometry here is not exactly the same as in \citet{Li2016}. In the SWH case, the SBL has the effect of further slowing down the speed, bringing to the formation of almost-zero velocity regions within the canopy. Differently, in the SLH case the SBL exerts a much lower reduction on the mean velocity field. It can be argued that since local heating and stable approaching flow have opposite effects on the mean velocity field, in this particular case the local heating overcomes the incoming stability. On average, the velocities in the canopy are reduced by 17, 32 and just 3\% for the SNH, SWH and SLH cases, respectively, compared to the NBL cases.

The observed TKE fields normalised by the reference velocity are reported in Figure~\ref{fig:TKE1}. In all cases the largest values of TKE are found in the region between $z=H$ and $1.5H$, above the canopy. A logarithmic scale was deemed necessary in order to adequately discern also the smallest variations of turbulence in the canopy (the averaged profiles on the right side, however, are in linear axes). In the WH case, the main feature is the presence of an increasingly turbulent region close to the heated wall, with the turbulence peaking around the upper windward street-canyon corner and spreading upstream. \citet{Allegrini2013} found the maximum TKE values in the same region, attributing this to the fact that there the cold air enters the canyon hitting the warmer air, which is rising due to buoyancy at the windward wall. The longitudinally-averaged profile appears to grow almost linearly in the canopy. A similar trend was also found by \citet{Park2012}, despite the fact that they only presented profiles at the vertical centreline. In the LH case, the increment in TKE in the canyon is more limited and not located near the heated wall, but closer to the windward wall (as also pointed out by \citet{Allegrini2013}). The slight reduction of TKE above the canopy is likely not generated by the leeward wall heating, but rather from the way the model was cooled. In fact, in order to allow the wind tunnel to remotely change from neutral to stable approaching flow, the cooling water used to refrigerate the unheated model surfaces was allowed to flow also in the rest of the wind tunnel floor. Since such water (to regulate the laboratory temperature) was set to 1$^\circ$C lower than the free stream one, the generated approaching flow presented a slightly positive temperature gradient, hence resulting in a very weak SBL, instead of a completely neutral one. This procedure was corrected for the WH case, which does not present this issue.

Finding the right scaling parameter for fluctuating quantities in this case is not trivial, as both local effects and incoming stratification may affect turbulence, especially in the canopy region. For this reason we have reported two sets of plots, with two different scaling parameters: (1) a reference wind speed (measured in the approaching flow at $z=2H$), which is the widely used way of normalising values in the literature and allows for a comparison with other studies, and (2) a reference TKE value (calculated using $u'$ and $w'$ measured in the incoming flow at $z=2H$), which takes into account the different levels of turbulence in the imposed boundary layers.

When scaled with the reference velocity, the stable stratification generates a strong and generalised reduction of TKE both above and inside the canopy, also in the presence of wall heating. This is estimated in an average decrease inside the canopy of 50, 46 and 30\%, respectively for the SNH, SWH and SLH cases, compared to the NBL cases. Despite the different local stratification, above $1.25H$ the TKE profiles collapse very well on each other in the SBL cases, meaning that the wall buoyancy-generated turbulence does not affect the SBL above. To be noted that the TKE reduction inside the canopy is not as large as for the approaching flow (see Figure~\ref{fig:BL}), for which the levels where almost four times lower after the application of the SBL. When TKE values are normalised by the incoming turbulence level, the stable stratification produces an increase estimated in 86, 121, and 158\%, respectively, for the SNH, SWH and SLH cases compared to the neutral approaching flow counterparts. This means that the influence of local obstacles and sources of heating on the local TKE field is, on average, stronger than the effect of the approaching flow, so that the reduction in the incoming flow turbulence levels do not match the decrease in TKE within the canyon.

\subsection{Temperature and heat flux} \label{Temperature and heat flux}

\begin{figure*}
	\centering
	\includegraphics[width=\textwidth]{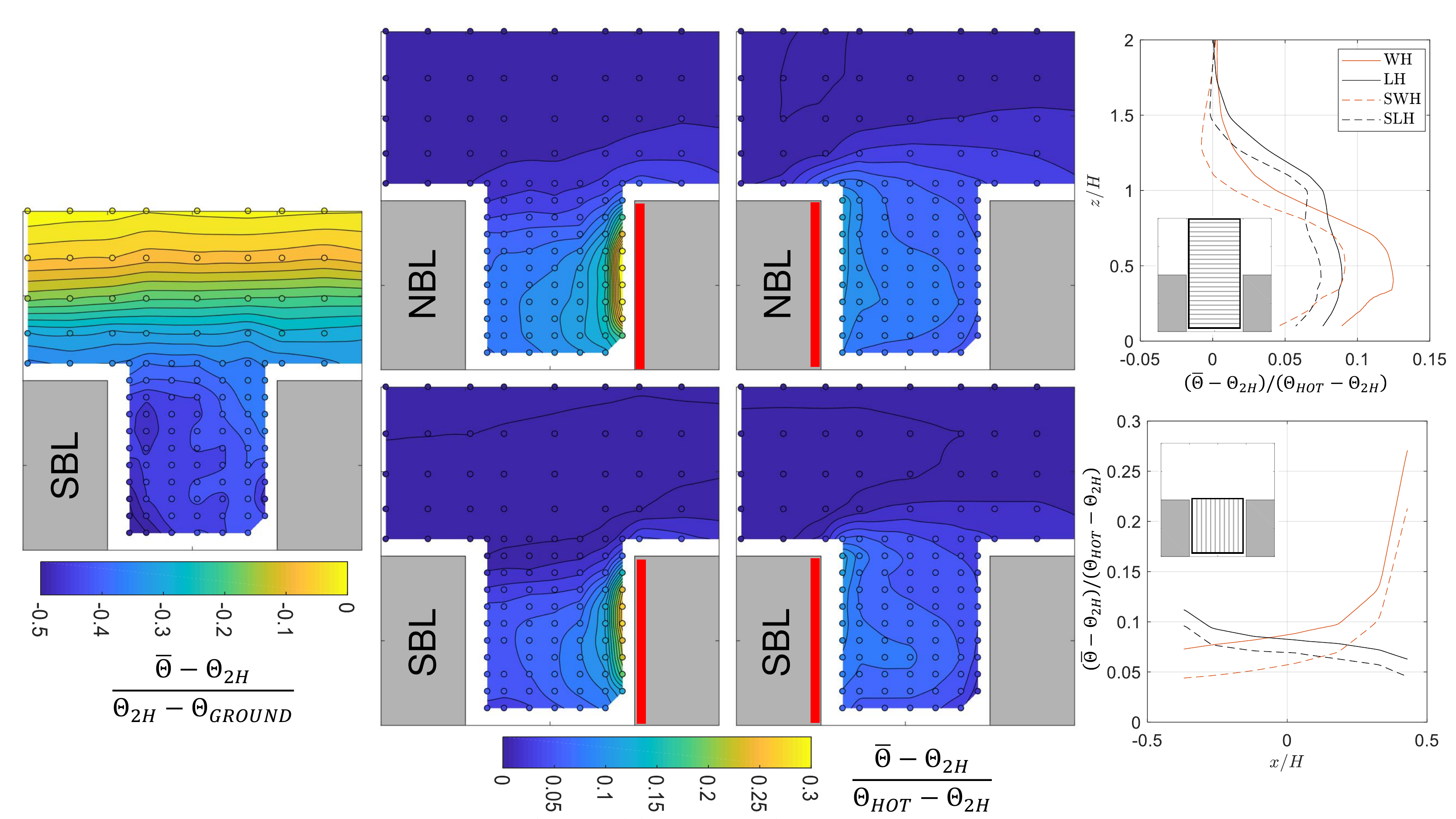}
	\caption{Contours of mean temperature. The red lines represent the heated surfaces in each case (SNH left, [S]WH centre, [S]LH right). The line plots on the right show the vertical profiles of longitudinally-averaged mean temperature (top) and the longitudinal profiles of vertically-averaged mean temperature (bottom); NBL = continuous lines, SBL = dashed lines.}
	\label{fig:T1}
\end{figure*}

Contour plots of mean temperature in the various cases are shown in Figure~\ref{fig:T1} (except the NH case where there is no temperature variation). Vertical and longitudinal profiles of longitudinally- and vertically-averaged mean temperature are also presented. For the SNH case the temperature is normalised as $(\overline{\Theta}-\Theta_{2H})/(\Theta_{2H}-\Theta_{GROUND})$. Above the canopy the temperature is clearly vertically stratified, while warmer air is observed sinking closer to the windward wall and raising colder along the leeward one, once being cooled by the floor. Thus, the stratification within the canopy appears to be directed horizontally across the canyon rather than vertically.

In the two wall-heated cases the temperature is normalised as $(\overline{\Theta}-\Theta_{2H})/(\Theta_{HOT}-\Theta_{2H})$. The warming effect appears to be confined near the heated wall. The WH case produces a larger increment in temperature compared to the LH case. However, because of the way the different instruments are mounted (see Section~\ref{Methodology}), the temperature measurement grid is 5~mm closer to the windward wall ($0.07H$ far) than to the leeward wall ($0.13H$). This contributes to the lower maximum temperatures observed for the LH and SLH cases. Keeping this in mind, it is noted that the averaged mean normalised temperature within the canopy is also higher for WH (0.104) than for LH (0.083). As pointed out by \citet{Cai2012}, they are representative of the warming efficiency of the heated wall on the canyon air. Above the canopy, though, the LH case presents larger temperatures compared to WH, meaning that the heating from the leeward wall is dispersed more in the upper region, as expected from the stronger mean vortex flow. The application of the incoming stable stratification appears to lower the temperature inside the canopy for both cases, without altering the shape of the longitudinally- and vertically-averaged profiles. It should be stressed that, due to the temperature gradient extending up to the boundary layer top, in the SBL cases the choice of a higher reference height for the temperature would affect the normalised temperature values, while for the NBL cases the air temperature above $2H$ is constant. Having this in mind, the averaged mean temperature within the canopy for the SWH and SLH cases are found to be 0.072 and 0.067, respectively, closer to each other compared to the two NBL cases.

\begin{figure*}
	\centering
	\includegraphics[width=\textwidth]{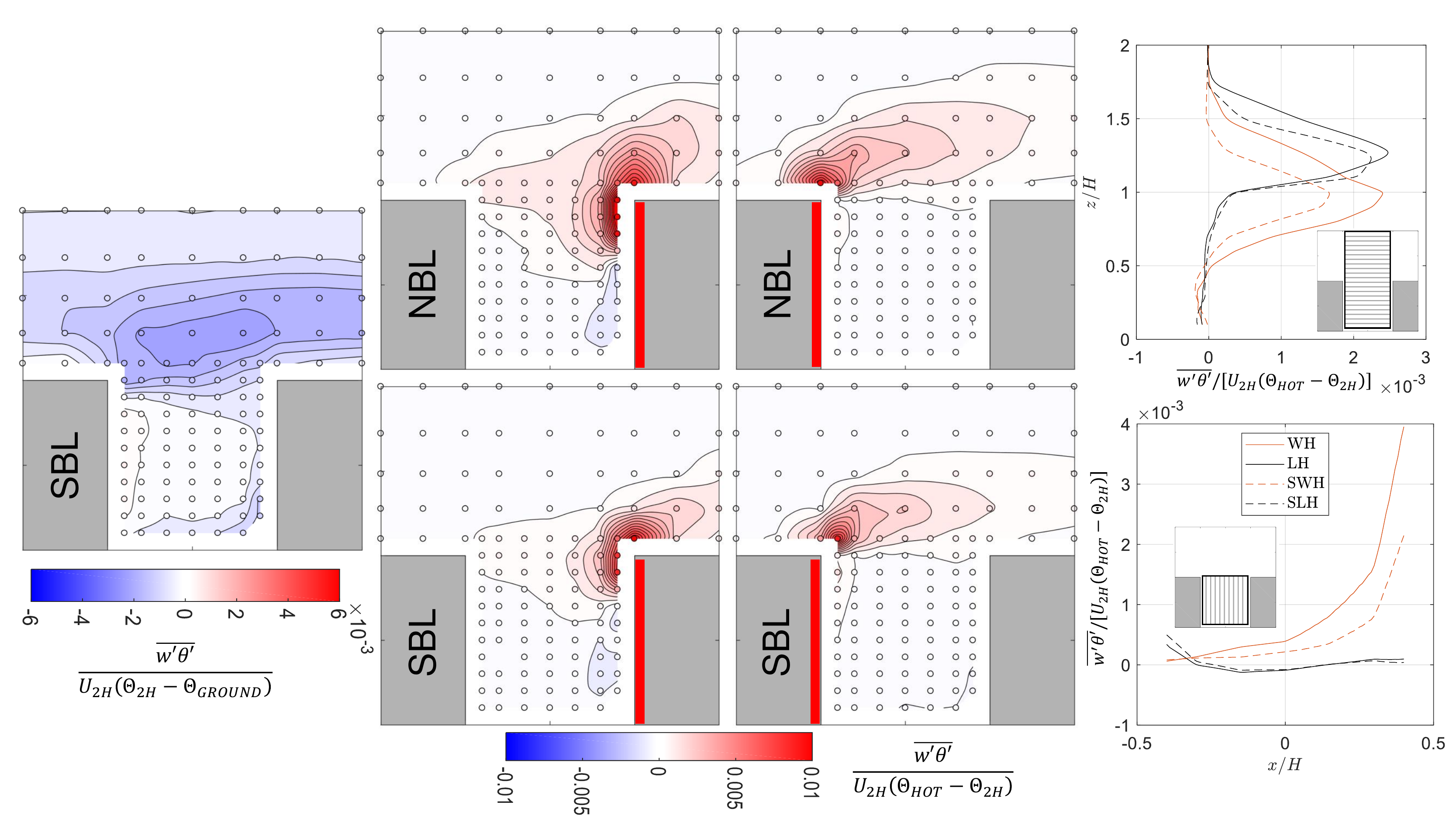}
	\caption{Contours of turbulent vertical heat flux. The red lines represent the heated surfaces in each case (SNH left, [S]WH centre, [S]LH right). The line plots on the right show the vertical profiles of longitudinally-averaged turbulent vertical heat flux (top) and the longitudinal profile of vertically-averaged turbulent vertical heat flux (bottom); NBL = continuous lines, SBL = dashed lines.}
	\label{fig:wt1}
\end{figure*}

Figure~\ref{fig:wt1} reports the graphs for the turbulent vertical heat flux. In the SNH case the flux is mainly negative, as expected for a SBL without a local source of heating. The maximum region is found in the shear layer immediately above the canopy, where the colder air raising from the street canyon faces the warmer upper region air. The heat flux in the canopy is larger closer to the exchange surfaces, while a region of slightly positive vertical heat flux is found closer to the leeward wall. Since only the floor surface is cooled, while the building walls are left passive, the colder air raises up facing the slightly warmer leeward wall, which in turns gives rise to the positive heat flux.

The heat flux field is obviously very dependent upon which surface is heated. The LH case is the one which affects less the heat flux distribution within the canopy, since the heated air is immediately released above the canopy and only a small part is re-entrained inside, although this point will be better analyse later through the the quadrant analysis. The flux peaks at the top of the leeward wall and spreads downstream over the canopy in the region of high shear. On the other hand, the WH case affects more the upper half of the canopy, with the heat flux peaking at the windward wall upper corner. Another feature is the presence of a slightly-positive flux region spreading up to the upper leeward building corner. Such flux is likely generated by the hot air trapped into the main vortex. Finally, a region of relatively strong negative heat flux is observed in the lower half of the canopy closer to the windward wall. Nevertheless, the longitudinally-averaged profiles display how, on average, the vertical flux is slightly negative in the lower half of the canopy for all the cases. They also highlight a vertical heat flux maximum for the LH at $1.25H$, moved down to $1H$ for the WH case. It is interesting to note that a similar location for the two maxima (even though only the profile along the centreline was shown) was also found by \citet{Park2012}. The application of the incoming SBL does not significantly modify the above analysis, but it contributes mainly to reducing the positive heat flux. The only exception is for SLH close to the leeward wall, where the SBL intensifies the positive heat flux. This is due to the fact that the cooling action of the windward wall and the floor reduces the temperature of the air approaching the heated wall, thus increasing the $\Delta\Theta$, and in turns the heat exchange.

\subsection{Pollutant concentration field} \label{Pollutant concentrations}
\begin{figure*}
	\centering
	\includegraphics[width=\textwidth]{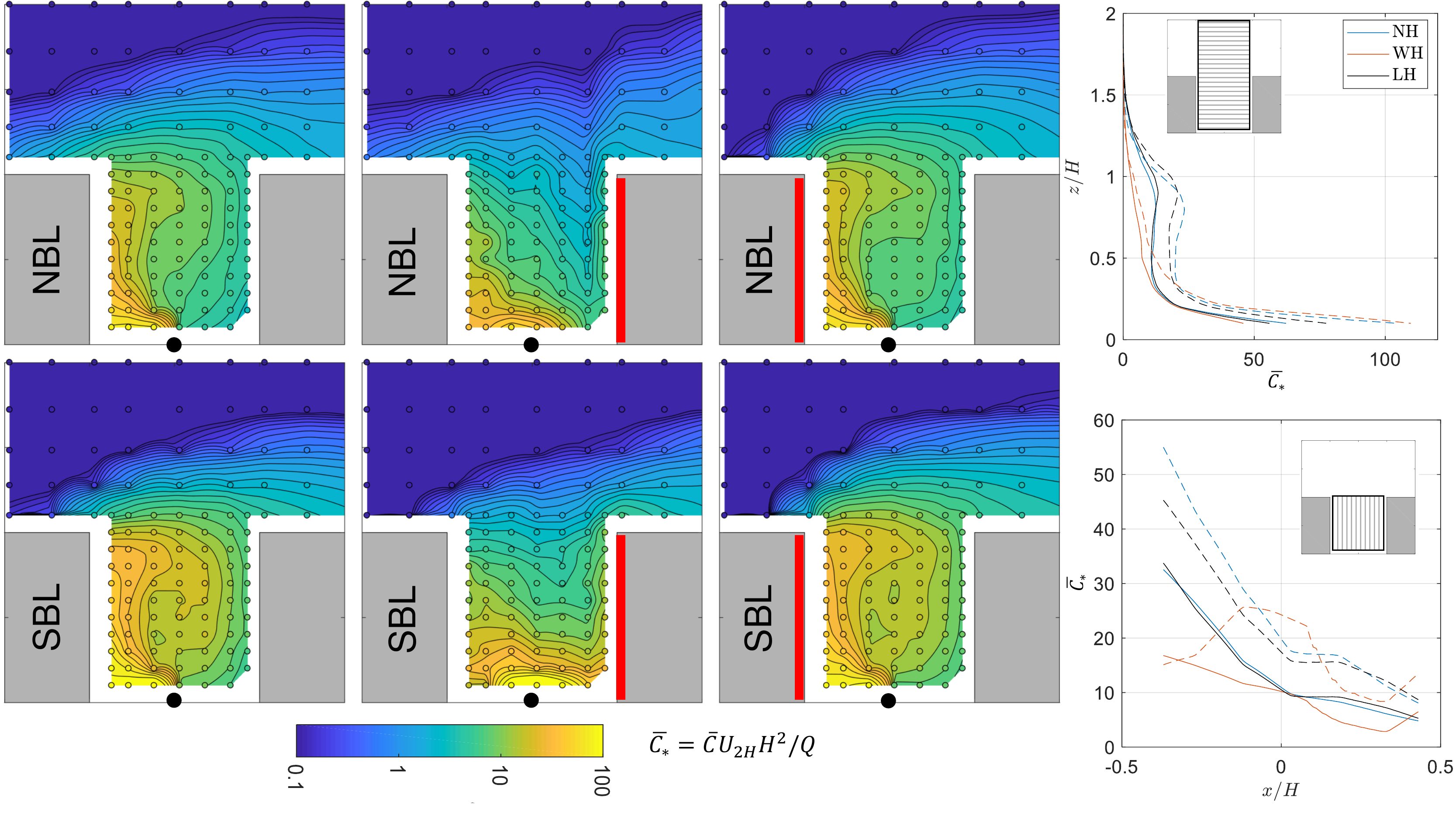}
	\caption{Contours of mean concentration. The red lines represent the heated surfaces in each case ([S]NH left, [S]WH centre, [S]LH right), the black circles represent the pollutant source. The line plots on the right show the vertical profiles of longitudinally-averaged mean concentration (top) and the longitudinal profiles of vertically-averaged mean concentration (bottom); NBL = continuous lines, SBL = dashed lines.}
	\label{fig:C1}
\end{figure*}

In this section we analyse the concentration field derived from releasing a passive tracer from a ground point source. Figure~\ref{fig:C1} shows the mean normalised concentration field in the cross-section for the six cases investigated in both logarithmic (contour plots on the left) and linear (averaged profiles on the right) scale. The concentration is normalised as $\overline{C_\ast} = \overline{C}U_{2H}H^2/Q$ where $Q$ is the pollutant tracer flow rate from the source. The isothermal case is characterised by a large concentration region upstream the source rising along the leeward wall up to the street canyon top, where some pollutant is re-entrained inside the canopy while other is carried downstream by the mean flow. In the WH case the pollutant transport by means of the main vortex is weakened by the action of the buoyancy force. Moreover, concentration values are increased downstream the source closer to the ground and along the windward wall, the latter due to pollutant up-drafts. The concentration pattern is very similar to what found by \citet{Cai2012a}, who simulated a scalar release from the entire street-canyon floor surface with windward wall and roof heating. For the LH case no significant differences are found in the cross-section compared to the NH case, despite the strengthened main vortex.

\begin{figure}
	\centering
	\includegraphics[width=\linewidth]{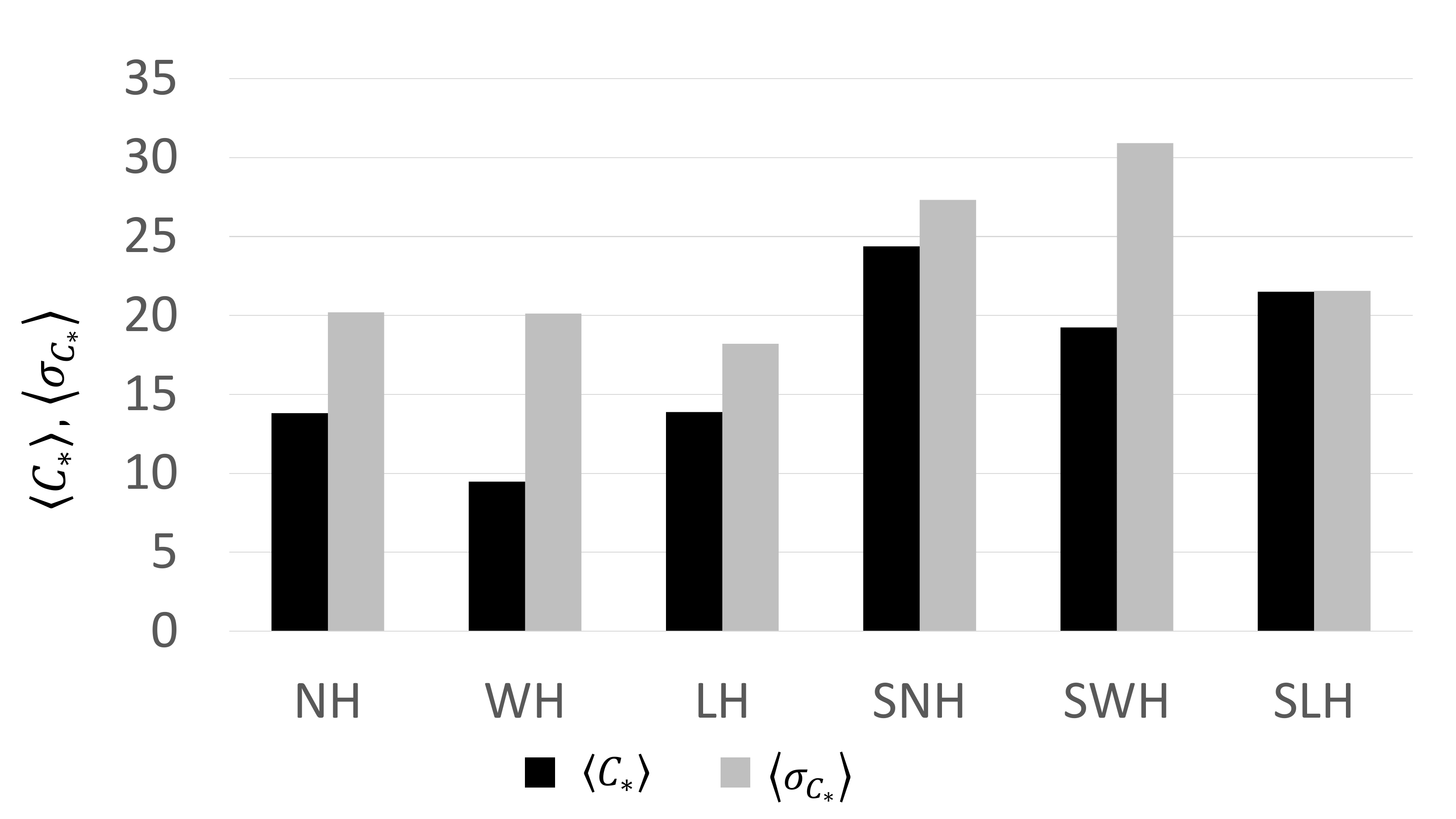}
	\caption{Normalised canyon cross-section averaged concentrations $\left<C_*\right>$ and normalised canyon cross-section averaged standard deviations of concentration fluctuation $\left<\sigma_{C_*}\right>$.}
	\label{fig:istC}
\end{figure}

The application of the incoming SBL creates a generalised increase of concentration inside the canopy, well summarised by the histogram in Figure~\ref{fig:istC}, which reports the values of normalised canyon cross-section averaged concentrations. For SNH the value is increased by about 75\% compared to the NH case. Such increment is very close to what found by \citet{Li2016} for a line source with a similar level of stratification. An even larger increment of concentration is experienced by the SWH case, which has a level of pollutant within the canopy that is double compared to the NBL counterpart. Such strong increase is concentrated mostly in the lower half of the canopy, thus more significant at pedestrian level. The increment for the SLH case is more modest, with a 55\% increase. Looking at the longitudinal profiles of vertically-averaged concentration, it is possible to observe how, while for the NH and LH case the high level of pollutant close to the leeward wall is even increased by the SBL, for the WH and SWH it is consistently lower. In the latter, the region of larger concentration is moved towards the centre of the canyon, driven by the velocity stagnation region which determines a large level of concentration immediately after the source release.

The standard deviation of the pollutant fluctuations averaged in the cross-section is reported in Figure~\ref{fig:istC}. For all cases the standard deviation is found to be larger than the mean value, often due to large (but quite sporadic) peaks in the signal (causing also a large positive skewness). This is particularly true for the WH case, where it is twice as large as the mean concentration within the canyon. The SBL has the effect of increasing the pollutant fluctuations, but less than the mean concentration, so that for the SLH case they have roughly the same value. \citet{Cai2012a} also reported fluctuations larger for the windward-heated case compared to the leeward-heated, but not exceeding 50\% of the mean concentration within the canopy. The larger value in this case can be explained by the choice of a point source instead of a surface release.

\subsection{Pollutant fluxes} \label{Pollutant fluxes}
The vertical pollutant fluxes are here considered. They can be divided into turbulent component ($\overline{w'_\ast c'_\ast}$), and mean ($\overline{W}_\ast \overline{C}_\ast$), while total fluxes can be given by the sum of the previous two ($\overline{w'_\ast c'_\ast} + \overline{W}_\ast \overline{C}_\ast$), where $W_\ast$ represents the vertical velocity normalised by $U_{2H}$.

\begin{figure*}
	\centering
	\includegraphics[width=\textwidth]{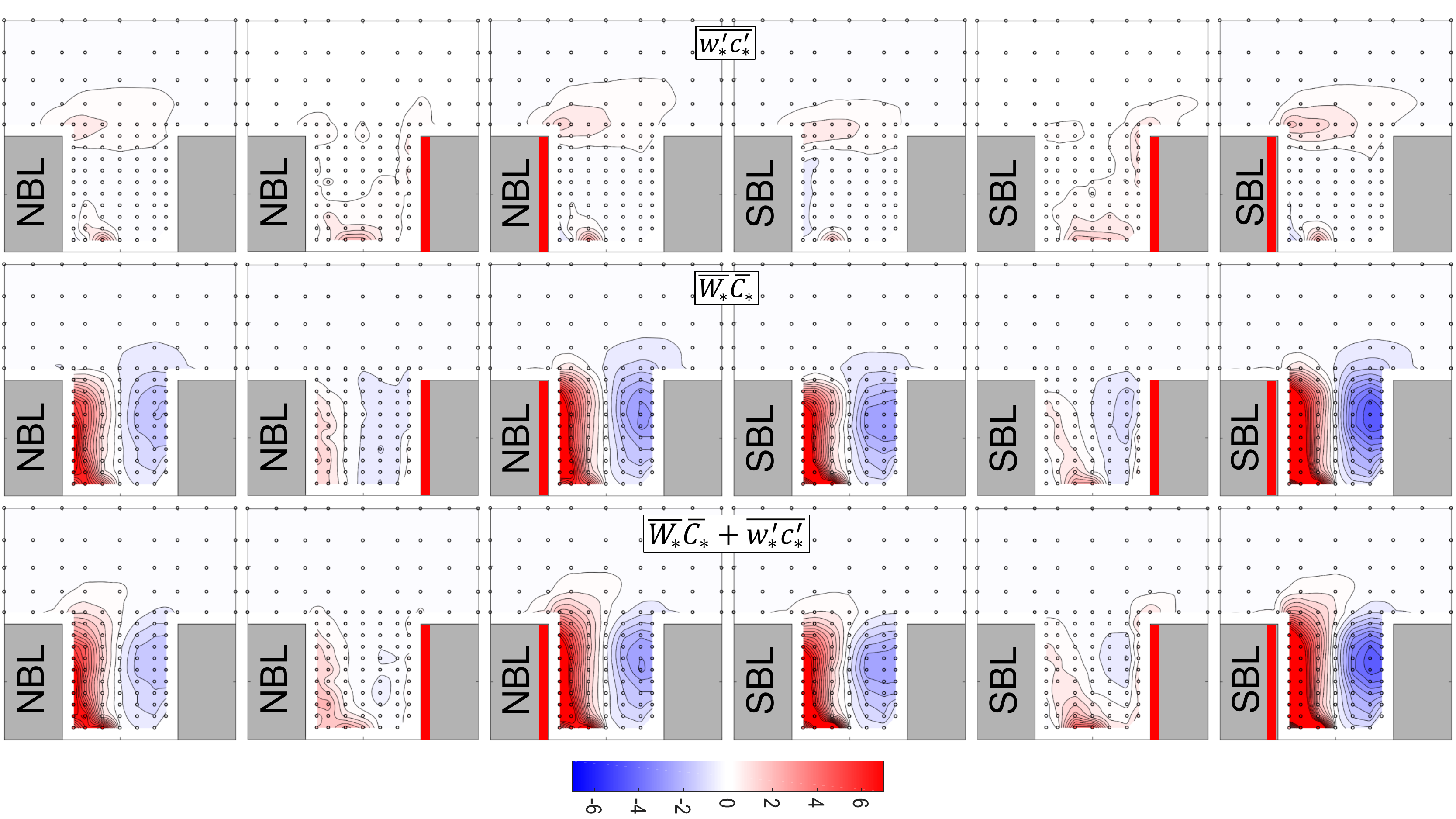}
	\caption{Contours of normalised vertical turbulent, mean and total pollutant flux. Velocities are normalised as $(w',\overline{W})_\ast=(w',\overline{W})/U_{2H}$, while concentrations as $(c',\overline{C})_\ast=(c',\overline{C})U_{2H}H^2/Q$.}
	\label{fig:wc1}
\end{figure*}

Figure~\ref{fig:wc1} shows the contours of the pollutant fluxes in the cross-section. For the isothermal case $\overline{w'_\ast c'_\ast}$ is only appreciable close to the source and at roof level (where it assumes positive values), while inside the canopy the mean flux controls the vertical pollutant exchange (with positive flux in the upstream half and negative in the downstream region of the street canyon, according to the mean vortex pattern). This result is in line with what observed by \citet{Carpentieri2012,Carpentieri2018} for more complex geometries. The LH case presents a similar trend, with a larger mean flux due to the increment in the mean velocity field. Differently, in the WH case turbulent fluxes are comparable to mean fluxes inside the canopy, due to the weakened mean flow. The negative total flux in the downstream half of the canyon almost disappears, since positive turbulent and negative mean flux counterbalance each other. A slightly-positive flux region is observed very close the windward wall, due to updrafts caused by the heated wall.

The application of the incoming stable stratification was found to have small effects on the turbulent pollutant fluxes, which are only slightly altered. In particular, a region of negative flux appears close to the leeward wall for the SNH case, which opposes the pollutant ventilation (as also observed by \citet{Li2016}). On the other hand, the large increment of concentration in the canopy almost everywhere overtakes the reduction in the mean velocity, hence the mean flux appears increased for all the cases. This is particularly true for the SLH case, where the velocity reduction was just 1\% (see section~\ref{Sec:Flow and turbulence}). In the SWH case, the positive flux region close to the heated wall appears strengthened by the SBL.

It must be stressed that since the pollutant release is not bi-dimensional, the vertical flux may be influenced by a variation in the lateral dispersion. On this aspect \citet{Sessa2018}, comparing the difference between point and linear source dispersion in stable atmosphere, pointed out that the effects of stratification on the first configuration are expected to be larger due to a reduced lateral spreading. On the other hand, \citet{Marucci2018a} did not observe significant variations of plume lateral dispersion  from a point source in a rectangular array of buildings for similar levels of stable stratification. It appears, then, that this aspect deserves further investigation.

\subsection{Exchange rates of pollutant and air}
\citet{Liu2005} introduced two useful parameters for evaluating canopy ventilation: the pollutant exchange rate (PCH) and the air exchange rate (ACH), computed by integrating the instantaneous vertical pollutant flux and vertical velocity, respectively, along the street canyon width $W$ at roof level.

\begin{align}
	PCH(t) &= \int_{W}w(t)c(t)dx \\
	ACH(t) &= \int_{W}w(t)dx
\end{align}

\begin{figure}
	\centering
	\includegraphics[width=\linewidth]{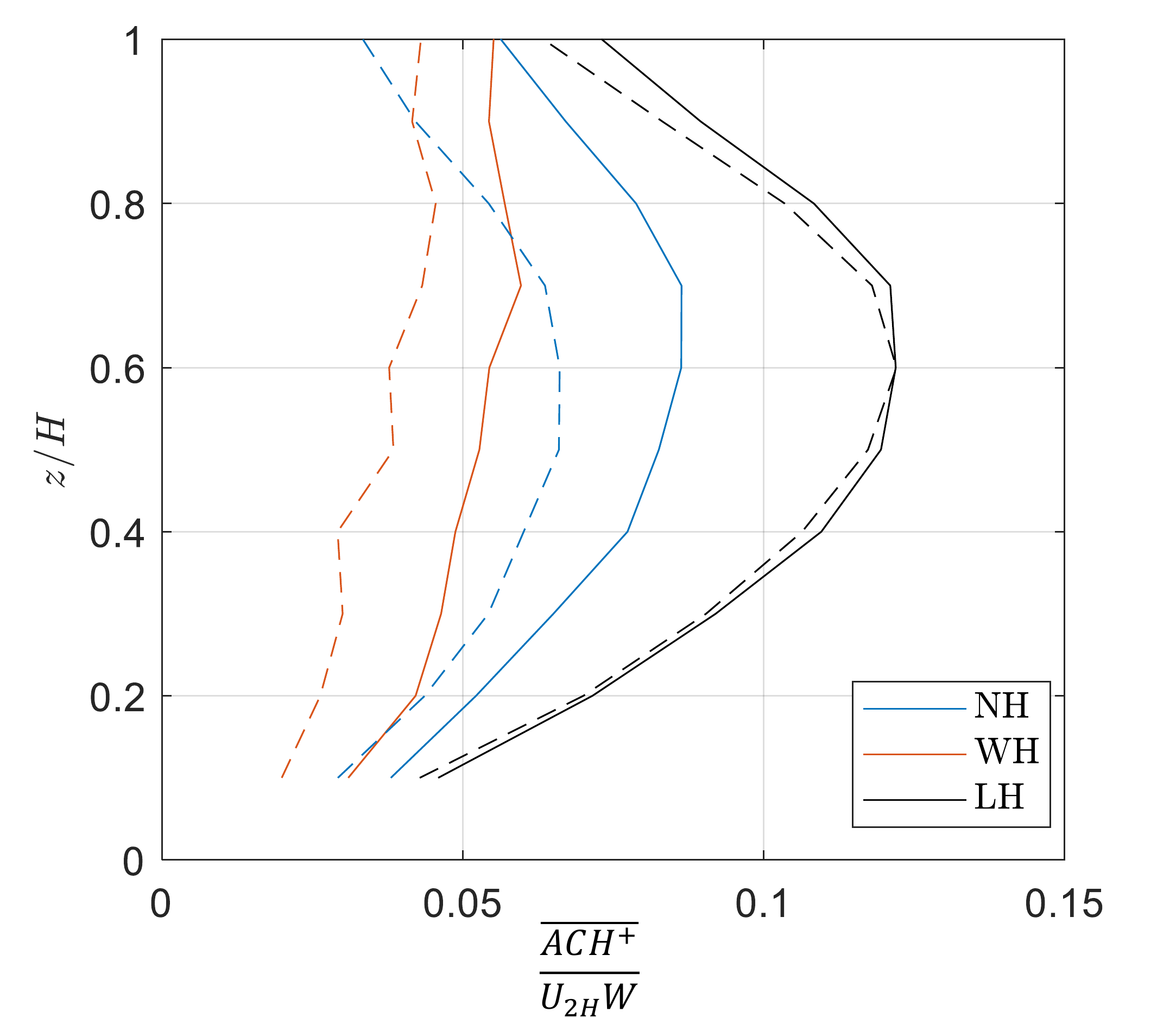}
	\caption{Vertical profiles of normalised $\overline{ACH^+}$. Continuous lines represent NBL data while dashed lines are SBL cases.}
	\label{fig:ACH}
\end{figure}

\begin{figure}
	\centering
	\includegraphics[width=\linewidth]{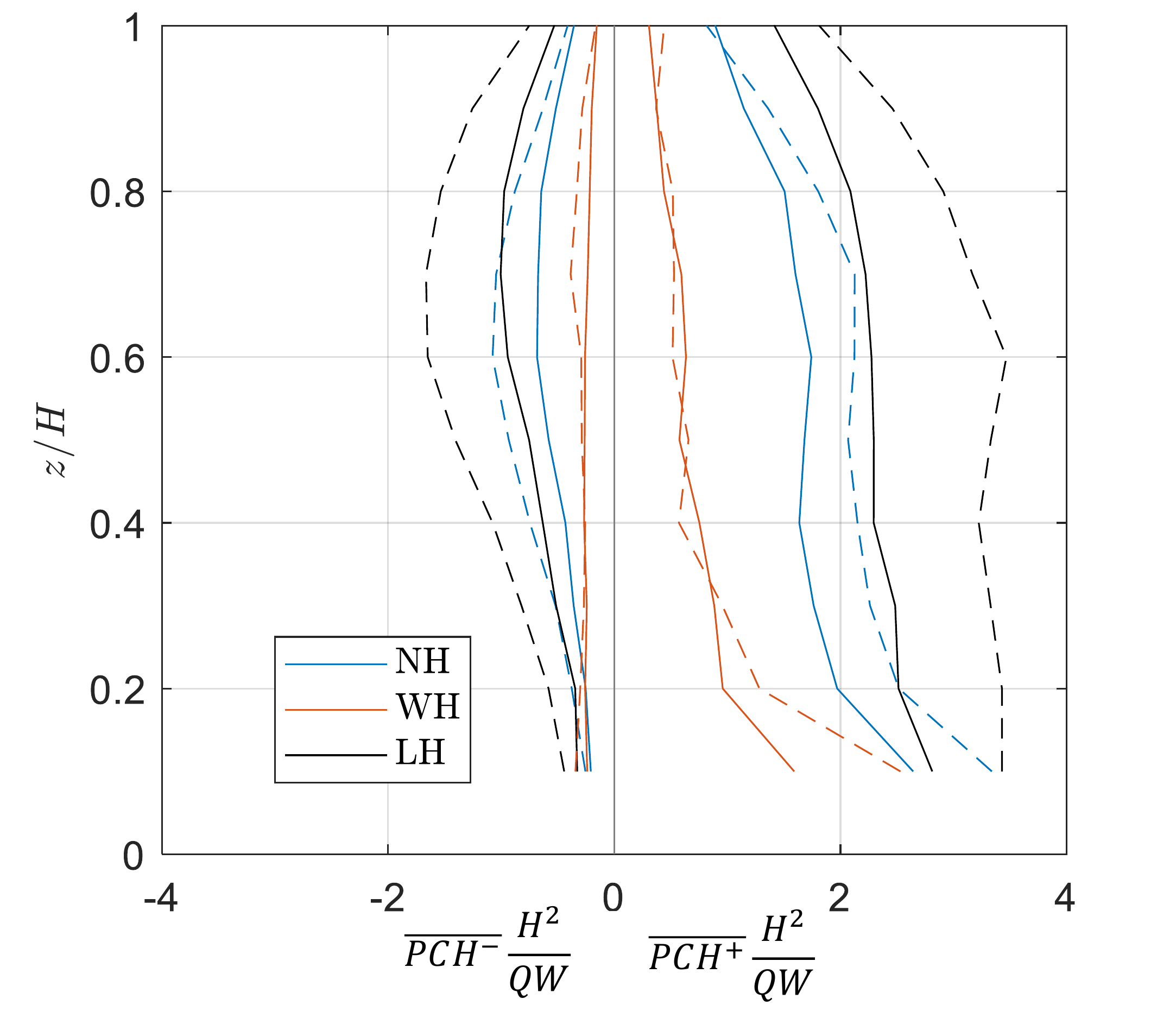}
	\caption{Vertical profiles of normalised $\overline{PCH^+}$ (on the right quadrant) and $\overline{PCH^-}$ (on the left). Continuous lines represent NBL data while dashed lines are SBL cases.}
	\label{fig:PCH}
\end{figure}

\noindent Their computation, though, requires the knowledge of instantaneous velocity and concentration fields in the whole integration domain, while in our case they were measured simultaneously only at single points. Despite this, we assume that the time-averaged rates ($\overline{PCH}$ and $\overline{ACH}$) can still be computed as

\begin{align}
	\overline{PCH} &= \overline{\int_{W}w(t)c(t)dx}=\int_{W}\overline{w(t)c(t)}dx \\
	\overline{ACH} &= \overline{\int_{W}w(t)dx}=\int_{W}\overline{w(t)}dx
\end{align}

\noindent providing that the measuring time is long enough to get statistically representative samples. The two rates can then be decomposed in $\overline{PCH^+}$, $\overline{ACH^+}$ and $\overline{PCH^-}$, $\overline{ACH^-}$ considering, respectively, only either positive or negative instantaneous velocity samples, while the others are alternatively imposed equal to zero. The positive rates represent the removal of pollutant/air from the street canyon, while the negative ones the pollutant/air re-entrainment into the cavity. It should be noted that air exchange rates at the canyon top correspond to the actual pollutant removal only by assuming well-mixed conditions within the canopy. However, particularly for a point source, this assumption is not well satisfied. For this reason, to get a better insight of the vertical ventilation, the exchange rates are computed at different heights in the canyon \cite{Garau2018}, as displayed in Figures~\ref{fig:ACH} and \ref{fig:PCH}.

In the isothermal case, $\overline{ACH^+}$ presents a maximum approximately at the height of the main vortex centre (as also found by \citet{Garau2018}) followed by a decrease up to the canyon top. The LH case shows a similar trend, but with amplified values due to the larger velocity magnitudes. On the other hand, in the WH case $\overline{ACH^+}$ almost monotonically increases with height, but with lower values compared to the other case. The application of the incoming stable stratification has the general effect of decreasing the exchange rate, following the reduction in the mean and fluctuating velocities discussed in section~\ref{Pollutant concentrations}. The observed decrease in the exchange rate is rather limited for the LH case, for which the stable stratification had only a small impact on the mean flow.

$\overline{PCH^+}$ presents a different trend, namely a reduction with  height thanks to the larger values of concentration in the bottom region. Despite this, the three local heating configurations are still organised with WH, NH and LH in growing order of exchange rate values. In this case, the effect of stable stratification is interestingly seen to produce opposite effects compared to $\overline{ACH^+}$. As a matter of facts, on average $\overline{PCH^+}$ is increasing within the canopy, especially for SLH, while the air exchange rate did not show a significant modification in that case. On the other hand, the SWH case does not show significant variations from WH. This discrepancy might be up to the fact that concentrations in SBL were found to increase more than the velocity reduction, hence resulting in an increase of $\overline{PCH^+}$ values. The effect is similar to what we observed in vertical pollutant fluxes (section~\ref{Pollutant fluxes}). Finally, $\overline{PCH^+}$ at roof level are found to be approximately twice as large as $\overline{PCH^-}$, confirming the results by \citet{Liu2005} and \citet{DiBernardino2018}, despite the different type of source.

\subsection{Quadrant analysis}
\begin{figure}
	\centering
	\includegraphics[width=\linewidth]{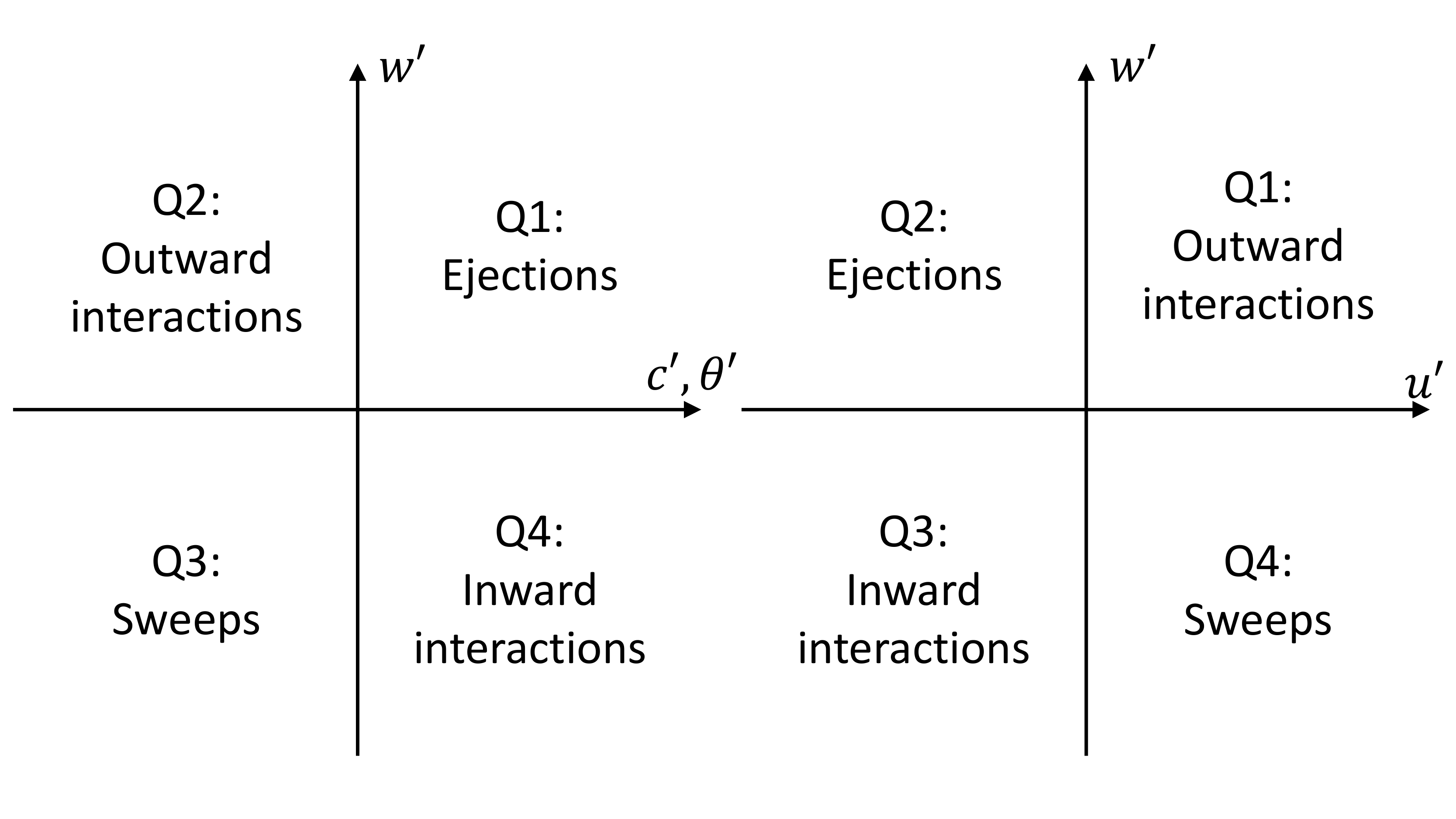}
	\caption{Scheme of quadrant division of the events for the vertical turbulent pollutant and heat flux (on the left), momentum flux (on the right).}
	\label{fig:QuadrantScheme}
\end{figure}

In this section we analyse the turbulence structure within and around the street canyon model by means of a quadrant analysis \cite{Wallace1972}, in which the fluctuations of two quantities at single locations are decomposed into four quadrants. Here three couples of parameters have been considered, namely the interactions between the vertical velocity fluctuations and the fluctuations of concentration, temperature and streamwise velocity. Various terminology has been employed in the literature to identify the events associated with the different quadrants. In the paper we adopt the terminology described in Figure~\ref{fig:QuadrantScheme}. Events characterised by a positive fluctuation of both vertical velocity and concentration or temperature are called ``ejections'' and represent the rise of more polluted/warmer air. On the other hand, negative fluctuations of both quantities are called ``sweeps'', representing the sink of cleaner/colder air. Both the events contribute positively to cleaning/cooling the air inside the canopy. Differently, a positive (or negative) fluctuation of vertical velocity coupled with a negative (or positive) fluctuation of concentration/temperature represents the rise of cleaner/colder air or the sink of more polluted/warmer air, hence contributing negatively to the ventilation within the street. Ejections and sweeps are often referred to as ``organised motions'' while inward and outward interactions as ``unorganised motions''.  As far as the momentum flux is concerned, the same terminology is adopted, but the phenomena are localised in different quadrants, according to Figure~\ref{fig:QuadrantScheme}.

\begin{figure*}
	\centering
	\includegraphics[width=0.9\textwidth]{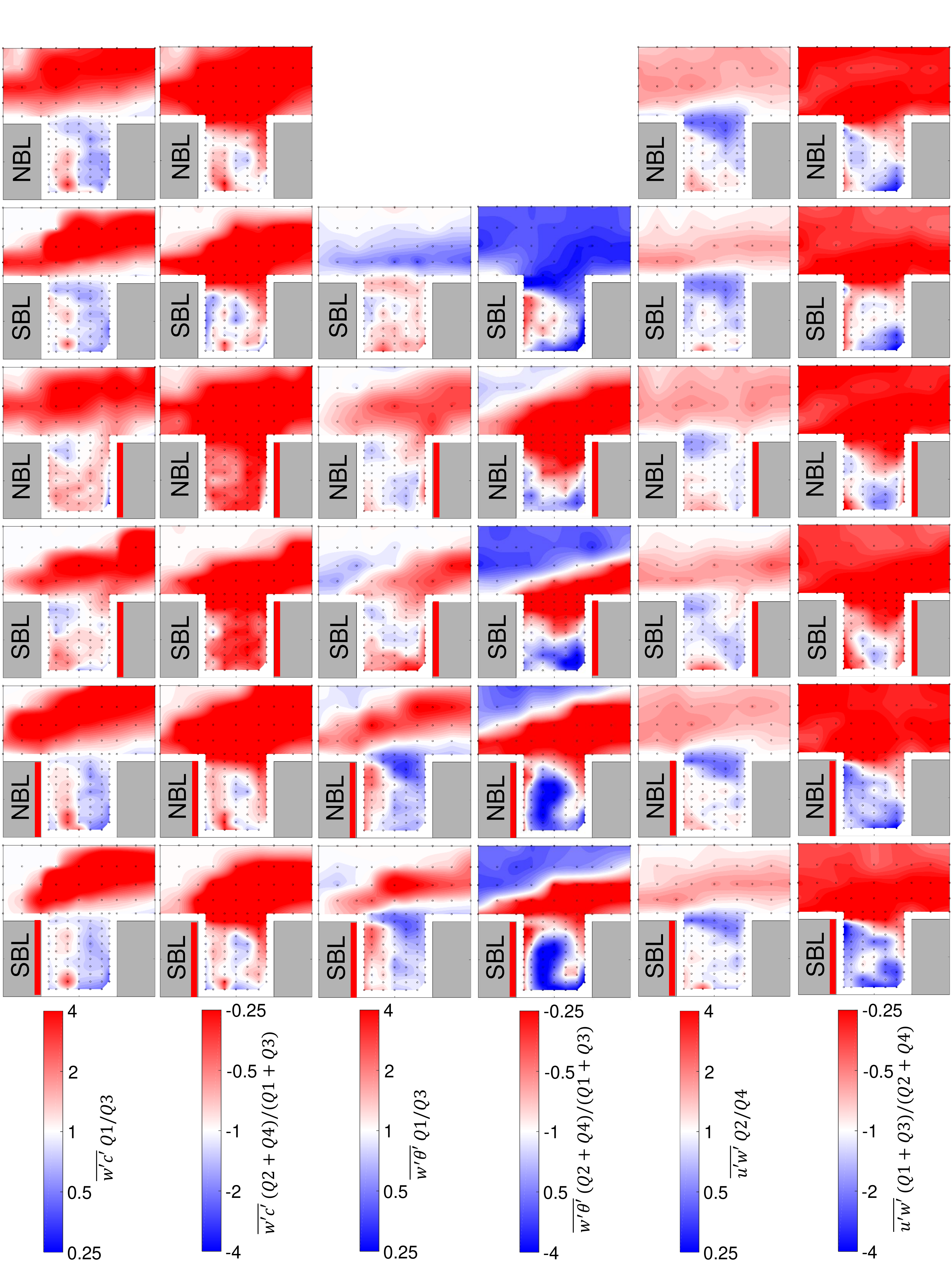}
	\caption{Ratio of ejection vs sweep (odd columns) and unorganised vs organised motion (even columns) contributions to the vertical turbulent pollutant, heat and momentum fluxes.}
	\label{fig:Quadrants}
\end{figure*}

Figure~\ref{fig:Quadrants} summarises the analysis for the investigated quantities by means of the ratio of ejections over sweeps as well as unorganised over organised motions. Such a visualisation is very compact and convenient, but it does not allow to distinguish the contribution of the inward from the outward interactions. When necessary, then, salient differences will be highlighted in the following description.
Moreover, special care should be taken in observing the graphs, since a large value of the ratio might result from a small numerator divided by an extremely small denominator. Nevertheless, in the following comments the predominance of a component on the other is highlighted only when effectively corresponding to a meaningful and genuine difference of magnitude of the component values (by looking at the data for each quadrant). Finally, it should also be noted that the quadrant analysis is meaningful only in case the turbulent contribution surpasses the mean flow. Hence, it is particularly significant for the [S]WH case but less for the other cases, characterised by a stronger mean flow. For completeness, though, all the cases are reported here.

With a neutral approaching flow, ejections and sweeps dominate the momentum transport above the canopy ($z/H>1.15$) with, respectively, 50\% and 30\% of the total contributions, as also found by \citet{Cheng2011a}. However, at roof level sweeps are found predominant over ejections, with almost inverted percentages. When the windward wall is heated, sweeps are reduced and ejections are predominant close to the heated corner. An increment of the ejections at roof level was also observed by \cite{Park2012}. The LH case is characterised by increased outward interactions closer to the heated wall. Inward interactions, on the other hand, are always accentuated on the lower downstream corner, apart for the windward heated cases, for which their peak is moved towards the canyon centre. When the incoming stable stratification is introduced, large ejections above the canopy are confined in the region between $z/H = 1.25$ and $1.5$, while above it they become comparable with sweeps. Unorganised motions within the canopy are also reinforced in the SLH case.

As far as the heat flux is concerned, ejections dominate in the WH case above the canopy, while at roof level sweeps are also determinant on the leeward wall side. In the lower half of the canopy the heat flux is negative (predominant unorganised motions) due to the cooling from the refrigerated ground and leeward wall. In the LH case, the effect of wall heating is barely seen in the canopy, as already pointed out in section~\ref{Temperature and heat flux}. As a matter of facts, the heat flux is mostly negative, with the exception of a strong sweep region of fresher air at roof level near the windward wall and, of course, ejections very close to the leeward heated wall. Above the canopy ejections of warm air departs from the upstream wall corner. In stable stratification (SNH case), unorganised motions are predominant as outward interactions above the canopy and inward interactions at roof level and along the windward wall. Positive heat flux in the form of sweep is only found close to the leeward wall. The main effect of the stable approaching flow in the wall heated cases is in confining the ejections of hot air closer to the canopy. Moreover, in SWH the stagnant region at the bottom of the canopy is controlled by inward interactions.

The turbulent pollutant flux in the isothermal case was found comparable with the mean only close to the source and at roof level (see section~\ref{Pollutant fluxes}). In the first location ejections are predominant, while at roof level and closer to the windward wall sweeps of cleaner air play an important role, in accordance with the findings by \citet{Cheng2011a} and \citet{Li2016}. The application of the stable stratification has the effect of reducing the ejections closer to the source and at the same time strengthening inward interactions in the upper left region (as also shown by \citet{Li2016}).
In the WH case the turbulent structure appears widely modified, with ejections controlling the turbulent transport everywhere except on the upstream side at roof level, where sweeps play an important role as well. Inward interactions in the canopy are extensively reduced, changing from 20\% down to 9\% of the total contributions. Conversely, the LH case does not present any significant modification in turbulent pollutant transport compared to the isothermal case. The incoming SBL in the SWH case has the effect of slightly enhancing sweeps, while in SLH the main modification is the reduction of the ejection closer to the heated wall.

\subsection{Limitations and future developments}
A 2D street canyon allows to identify and describe the various flow patterns by means of sampling only the central cross-section. On the other hand, more complex 3D geometries could greatly affect the local flow fields. In this regard, the wall heated street canyon model was designed with the buildings made of two identical parts in order to be able in the future to simulate also a case with an intersection. Experiments with a different street aspect ratio are also possible with the present set-up and would increase the dataset completeness. Moreover, the implementation of a linear source in place of the point source used here would be an important improvement in order to achieve bi-dimensionality in the plume pattern as well. Finally, further developments may come by considering a range of different local and incoming stratification levels.

\section{Conclusion}
\label{Conclusion}
An experimental campaign has been carried out, aiming to investigate buoyancy effects on flow and dispersion characteristics in a bi-dimensional isolated street canyon of unity aspect ratio. Both local heating, by means of heating the windward or the leeward canyon wall, and different approaching flow stratification (neutral and stable) have been considered.

As far as the mean velocity field is concerned, a single-vortex structure was observed in all cases, except when the windward wall was heated. In this case a counter-rotating vortex formed close to the heated wall, resulting in a reduction of the velocities within the canopy. Conversely, heating the leeward wall produced a considerable increment in the vortex speed. The incoming stable stratification was only found significant in the reduction of the velocities in the lower half of the canopy. In terms of turbulent kinetic energy, larger values were found above the canopy, while inside the street canyon the windward heated case produced the largest increment, in particular (but not only) close to the heated wall region. On the other hand, the leeward wall-heated case did not produce a significant increment of turbulence closer to the heated wall. Incoming stable stratification was found to produce a large and generalised reduction of turbulence both inside and above the canopy in all the cases when normalised by the reference velocity. The opposite when the approaching flow turbulence normalisation is applied.

Analysing heat exchange, the windward wall-heated case produces larger temperature increments within the canopy than the leeward case, for which the heat vacates immediately the canyon, as evidenced by the larger temperature and heat flux above the canopy. In any case, larger temperature increments are confined close to the heated walls. The stable stratification has the effect of lowering the temperature inside the canopy, as well as the positive vertical heat flux.

Tracer released from a ground level point source highlighted how the largest modifications in the plume cross-section can be expected when the windward wall is heated. In this case, breaking the updraft close to the leeward wall increases the pollutant level on the windward side. Leeward wall heating was not found to produce significant modifications on the plume shape and concentration levels. The application of the incoming stable stratification created a generalised increment of pollutant in the canopy, with concentrations up to twice as large. From the point of view of the vertical pollutant fluxes, the turbulent component was found comparable with the mean only close the source and at roof level. Differently in the windward-heated case, with the weakening of the main vortex, the two components are comparable with each other. The stable stratification does not affect considerably the turbulent exchange, but id does reinforces the mean.

Finally, a quadrant analysis was also performed on the vertical fluxes of momentum, heat and pollutant, in order to highlight the modifications in the turbulence structure caused by buoyancy effects.

These results highlight the importance of considering local and approaching flow stratification when dealing with urban ventilation and dispersion studies. The dataset produced can be valuable for validating CFD simulations on bi-dimensional street canyons. For this purpose, in future work a linear source could be employed to obtain a bi-dimensional plume, even though such implementation on actively cooled surfaces can be quite demanding. The study can be also further extended by considering more complex three-dimensional geometries, like urban intersections.

\section*{Acknowledgments}
This work was funded by the EPSRC (grant EP/P000029/1) and by the Department of Mechanical Engineering Sciences (University of Surrey). The authors confirm that all wind tunnel data are fully available without restriction from \href{https://doi.org/10.6084/m9.figshare.7804454}{https://doi.org/10.6084/m9.figshare.7804454}.

\bibliography{BuildingAndEnvironment}








\end{document}